\documentclass[aps, prd, twocolumn, superscriptaddress, footinbib, floatfix]{revtex4-2}
\usepackage{xcolor}
\usepackage{graphicx}
\usepackage{wrapfig}
\usepackage{amsmath}
\usepackage{amsfonts}
\usepackage{amssymb}
\usepackage{multirow}
\usepackage{slashed}
\usepackage{physics}
\usepackage{float}
\usepackage{bm}
\usepackage{soul}
\usepackage[colorlinks=true, linkcolor=blue, citecolor=blue, urlcolor=blue]{hyperref}

\usepackage[utf8]{inputenc}
\usepackage[normalem]{ulem}

\DeclareRobustCommand{\Eq}[1]{Eq.~\eqref{eq:#1}}

\DeclareRobustCommand{\fig}[1]{Fig.~\ref{fig:#1}}

\DeclareRobustCommand{\refcite}[1]{Ref.~\cite{#1}}

\newcommand\bets{\begin{table*}}
\newcommand\eets[1]{\label{tb:#1}\end{table*}}
\begin{document}
\widetext
\title{Lattice QCD Benchmark of Proton Helicity and Flavor-Dependent Unpolarized Transverse Momentum-Dependent Parton Distribution Functions at Physical Quark Masses}
\author{Dennis Bollweg}
\affiliation{Computing and Data Sciences Directorate, Brookhaven National Laboratory, Upton, New York 11973, USA}
\author{Xiang Gao}
\email{xgao@bnl.gov}
\affiliation{Physics Department, Brookhaven National Laboratory, Upton, New York 11973, USA}
\author{Swagato Mukherjee}
\affiliation{Physics Department, Brookhaven National Laboratory, Upton, New York 11973, USA}
\author{Yong Zhao}
\affiliation{Physics Division, Argonne National Laboratory, Lemont, Illinois 60439, USA}

\date{\today}

\begin{abstract}
We present the first lattice QCD calculations of the isovector helicity transverse momentum-dependent parton distribution function (TMDPDF) and the flavor-dependent unpolarized TMDPDFs for up and down quarks in the proton. Our computations utilize domain-wall fermion discretization with physical quark masses. Employing Coulomb-gauge-fixed quark correlation functions within the large-momentum effective theory framework, we access nonperturbative transverse quark separations $b_T$ up to approximately 1~fm, corresponding to transverse momenta as low as 200 MeV. Based on the quasi-TMD factorization theorem, we construct renormalization-group–invariant ratios that are equal to the corresponding light-cone TMDPDF ratios.
At moderate $x$, our results reveal that the isovector helicity and unpolarized TMDPDFs exhibit nearly identical transverse structure up to a normalization factor, and the unpolarized distributions display only mild flavor dependence. These findings not only support key trends observed in recent global analyses but also provide robust, nonperturbative constraints that can distinguish between different parametrizations. This work establishes a first-principles benchmark for TMDPDFs, offering valuable input for ongoing and future experimental efforts to map the proton’s three-dimensional structure.
\end{abstract}

\date{\today}
\maketitle

%%%%%%%%%%%%%%%%%%%%%%%%%%%%%%%%%%%%%%%%%%%%%%%%%%%%%%%%%%%%%%%%%%%%%%

\emph{Introduction.} Protons, as the stable constituents of atomic nuclei, form the basis of visible matter. Understanding their internal structure and dynamics from first principles of Quantum Chromodynamics (QCD), the theory of strong interactions, is a central goal of modern theoretical and experimental physics. This requires unraveling how the proton’s spin emerges from its quark and gluon constituents. While collinear parton distribution functions (PDFs) provide information on longitudinal momentum structure, a complete picture of proton spin demands access to its full three-dimensional momentum structure. Transverse momentum dependent PDFs (TMDPDFs) generalize PDFs by incorporating dependence on both the longitudinal momentum fraction $x$ and transverse momentum $k_T$. This makes TMDPDFs central to the scientific objectives of experiments at facilities such as Jefferson Lab~\cite{Dudek:2012vr, Accardi:2023chb}, the Large Hadron Collider (LHC)\cite{Collins:1984kg, Amoroso:2022eow}, and the upcoming Electron-Ion Collider (EIC)\cite{Boer:2011fh,Accardi:2012qut,AbdulKhalek:2022hcn,Abir:2023fpo}, which will explore the proton’s spin structure with unprecedented precision.

Among leading-twist TMDPDFs, the helicity distribution encodes the density of longitudinally polarized quarks in a longitudinally polarized proton, while retaining $k_T$ dependence. It plays a key role in resolving spin and orbital contributions of partons to the proton's total spin. The unpolarized TMDPDF, in contrast, describes intrinsic $k_T$ distributions irrespective of spin and serves as a baseline for spin-dependent observables. The flavor dependence of TMDPDFs is sensitive to valence–sea structure and isospin symmetry breaking, and is essential for interpreting transverse spin asymmetries and constraining global fits.

Phenomenological parametrizations of the unpolarized TMDPDF from global fits to semi-inclusive deep inelastic scattering and Drell-Yan data have made substantial progress in the past four decades~\cite{Davies:1984sp,Ladinsky:1993zn,Landry:2002ix,Konychev:2005iy,Sun:2014dqm,DAlesio:2014mrz,Bacchetta:2017gcc,Scimemi:2017etj,Bertone:2019nxa,Scimemi:2019cmh,Bacchetta:2019sam,Hautmann:2020cyp,Bury:2022czx,Bacchetta:2022awv,Isaacson:2023iui,Aslan:2024nqg,Vladimirov:2019bfa,Cerutti:2022lmb,Barry:2023qqh,Bacchetta:2025ara,Moos:2025sal}, including recent studies of its flavor dependence~\cite{Moos:2023yfa,Bacchetta:2024qre}. This approach is limited by sparse data in the kinematic region sensitive to nonperturbative physics and rely on model-dependent assumptions at low $k_T$ , where there are still tensions among different groups. In contrast, global fits of the proton helicity TMDPDF have only become available very recently~\cite{Yang:2024drd,Bacchetta:2024yzl}, but the results from the two groups show noticeable inconsistencies.

Lattice QCD is a first-principles approach that can calculate the spin- and flavor-resolved hadron structure in the nonperturbative regime. Although TMDPDFs are defined from time-dependent light-cone correlations and inaccessible on a Euclidean lattice, they can be computed within the large-momentum effective theory framework~\cite{Ji:2013dva,Ji:2014gla,Ji:2020ect} via the quasi-TMDPDFs~\cite{Ji:2014hxa,Ji:2018hvs,Ebert:2018gzl,Ebert:2019okf,Ji:2019sxk,Ji:2019ewn,Ebert:2020gxr,Vladimirov:2020ofp,Ji:2020jeb,Ji:2021znw,Ebert:2022fmh,Schindler:2022eva,Zhu:2022bja,Rodini:2022wic}. At large hadron momentum, the quasi-TMDPDFs are matched to their light-cone counterparts through an effective theory expansion~\cite{Ebert:2019okf,Ji:2019sxk,Ji:2019ewn,Ebert:2022fmh}, enabling access to observables such as the Collins-Soper (CS) kernel for TMD evolution~\cite{Shanahan:2020zxr,Shanahan:2021tst,Schlemmer:2021aij,Shu:2023cot,LatticeParton:2020uhz,Li:2021wvl,LPC:2022ibr,LatticePartonLPC:2023pdv,Avkhadiev:2023poz,Avkhadiev:2024mgd}, soft function~\cite{LatticeParton:2020uhz,Li:2021wvl,LatticePartonLPC:2023pdv}, and nucleon and pion TMDPDFs~\cite{LPC:2022zci,Chu:2023jia,Walter:2024nvq,LPC:2025spt}. 

Traditional lattice calculations of quasi-TMDPDFs use nonlocal quark bilinears with staple-shaped Wilson lines, which introduce linearly power divergence and substantial statistical noise, especially at large $b_T$ (or equivalently small $k_T$), thus limiting the precision. The recently developed Coulomb-gauge (CG) quasi-TMDPDF approach~\cite{Gao:2023lny,Zhao:2023ptv} addresses these issues by eliminating Wilson lines entirely, which not only reduces the statistical noise but also simplifies the renormalization. The CG method has enabled improved CS kernel calculations~\cite{Bollweg:2024zet,Mukherjee:2024xie} and the first lattice determination of the pion TMDPDF with the soft function~\cite{Bollweg:2025iol}, demonstrating its potential for advancing TMD studies.

In this Letter, we present the first lattice QCD calculation of the isovector helicity and flavor-dependent unpolarized TMDPDFs in the proton, taking advantage of the CG method. 
The calculation employs chiral-symmetry-preserving domain-wall fermion discretization and physical quark masses, and reaches the nonperturbative region up to $b_T \approx 1~\mathrm{fm}$ which corresponds to $k_T\sim 200$ MeV. By forming renormalization- and factorization-scale-independent ratios, we enable direct comparisons with phenomenological fits without requiring a separate determination of the soft function. Our results show a remarkable similarity between the isovector helicity and unpolarized TMDPDFs at moderate $x$, and reveal only mild flavor dependence in the unpolarized TMDPDFs—providing the first lattice benchmark in these channels. These findings offer valuable nonperturbative constraints for global analyses and establish a solid foundation for future lattice studies of the proton's 3D spin structure.

%%%%%%%%%%%%%%%%%%%%%%%%%%%%%%%%%%%%%%%%%%%%%%%%%%%%%%%%%%%%%%%%%%%%%%

\emph{Theoretical framework.} The bare CG quasi-TMD beam functions are defined as matrix elements of nonlocal quark bilinears with spatial separations in both the transverse ($b_T$) and longitudinal ($b_z$) directions, sandwiched between boosted hadron states,
\begin{align} \label{eq:OCG}
    \tilde{h}^{B}(b_T,b_z,P_z,\mu) =
    \langle \lambda;P_z|\mathcal{O}_\Gamma (b_T,b_z)|\lambda;P_z\rangle ,
\end{align}
with $\mathcal{O}_\Gamma (b_T, b_z) = \bar{\psi}(b_T,\frac{b_z}{2})\Gamma \psi(0,-\frac{b_z}{2})|_{\nabla\cdot\textbf{A}=0}$. Here, $\psi$ is the quark field, $\nabla\cdot\textbf{A}=0$ is the CG condition, $P_z$ is the hadron momentum, $\lambda$ the hadron polarization, and $\Gamma$ specifies the TMDPDF spin structure. This contrasts with the gauge-invariant (GI) quasi-TMD beam function, where non-local quark fields are connected by staple-shaped Wilson lines to preserve gauge invariance (see, e.g., \refcite{Ji:2014hxa}).

After appropriate renormalization of $\tilde{h}^{B}$---as explained below---and Fourier transform (FT) of the renormalized operator $\tilde{h}^{R}$ with respect to $b_z$, one obtains the quasi-TMD beam function $\tilde{f}(x,b_T,P_z,\mu)$, which is related to the light-cone TMDPDF $f(x,b_T,\zeta,\mu)$ via the factorization formula~\cite{Ebert:2019okf,Ji:2019sxk,Ji:2019ewn,Ebert:2022fmh,Zhao:2023ptv},
\begin{align}\label{eq:qTMDfac}
\begin{split}
	&\frac{\tilde{f}(x,b_T,P_z,\mu)}{\sqrt{S_r(b_T,\mu)}}=H(xP_z,\mu)\exp\left[{\frac{\gamma^{\overline{\rm MS}}(b_T,\mu)}{2}\ln\frac{(2xP_z)^2}{\zeta}}\right]\\
	&\times f(x,b_T,\zeta,\mu)+\mathcal{O}\left(\frac{\Lambda_{\rm QCD}}{xP_z},\frac{1}{b_T (xP_z)}\right),
\end{split}
\end{align}
with power corrections suppressed by $1/(xP_z)$~\cite{Liu:2023onm}, so that the factorization is valid in the moderate-$x$ region but is expected to break down near the endpoints. Here, $\mu$ and $\zeta$ are the factorization and rapidity scales. The evolution in $\zeta$ is governed by the CS kernel $\gamma^{\overline{\rm MS}}(b_T,\mu)$~\cite{Collins:1981uk,Collins:1984kg}. $H(xP_z,\mu)$ is a hard kernel obtained at one-loop~\cite{Zhao:2023ptv}, and $S_r(b_T,\mu)$ is the intrinsic soft factor~\cite{Ji:2019sxk}.

This factorization is independent of quark flavor and spin and does not involve gluon mixing~\cite{Ebert:2022fmh}. These features enable the construction of renormalization-group–invariant (RGI) ratios from quasi-TMD beam functions that are free from scheme and scale dependence. In particular, the cancellation of the soft factor, hard kernel, and scale evolution terms makes the RGI ratios identical to their light-cone counterparts to all orders in perturbation theory, up to power-suppressed corrections.

For example, the ratio of proton isovector helicity TMDPDF $g_{1L}^{\Delta u_+ - \Delta d_+}$ to isovector unpolarized TMDPDF $f_1^{u_v - d_v}$ can be expressed as
\begin{align} \label{eq:HeliUnpol} 
\begin{split}
    R^{u-d}_{g_{1L}/f_1}(x,b_T) \cdot g_A & \equiv
    \frac{g_{1L}^{\Delta u_+ - \Delta d_+}(x,b_T,\zeta,\mu)}{f_1^{u_v - d_v}(x,b_T,\zeta,\mu)} \\
    & = \frac{\tilde{g}_{1L}^{\Delta u_+ - \Delta d_+}(x,b_T,P_z,\mu)}{\tilde{f}_1^{u_v - d_v}(x,b_T,P_z,\mu)} ,
\end{split}    
\end{align}
where $g_A$ is the isovector axial charge of the proton and accounts for an overall normalization. Similarly, the RGI ratio of unpolarized up $u$- and down $d$-quark TMDPDFs of proton is
\begin{align}\label{eq:UnpolUD}
R_{f_1}^{u/d}(x,b_T) \equiv \frac{f_1^{u_v}(x,b_T,\zeta,\mu)}{f_1^{d_v}(x,b_T,\zeta,\mu)} 
= \frac{\tilde{f}_1^{u_v}(x,b_T,P_z,\mu)}{\tilde{f}_1^{d_v}(x,b_T,P_z,\mu)} .
\end{align}
Here, $q_v= q - \bar{q}$ corresponds to the valence quark contributions to the proton unpolarized TMDPDFs and $\Delta q_+=\Delta q + \Delta \bar{q}$ represents the combined contribution of quarks and antiquarks to the proton helicity TMDPDF, where $q=u$ or $d$.

\emph{Lattice QCD calculations.} The quasi-TMD beam function matrix elements are extracted from proton three-point correlation functions,
\begin{align}\label{eq:3pt}
\begin{split}
& C^{\mathrm{3pt}}_{\mathcal{P}, \Gamma}(t_s,\tau; b_T, b_z, P_z) = \sum_{\vec{y}, \vec{x}} e^{-i P_z (y_z - x_z)} \mathcal{P}^{\mathrm{3pt}}_{\alpha \beta} \\
& \times \langle N_\alpha (\vec{y}, t_s + t) \mathcal{O}_\Gamma (b_T, b_z, \tau + t) \overline{N}_\beta (\vec{x}, t) \rangle .
\end{split}
\end{align}
We use the proton interpolating field $N^s_\alpha(\vec{x}, t) = \varepsilon_{abc} u^s_{a\alpha}(\vec{x}, t) \left[u^s_b(\vec{x}, t)^T (\gamma_t \gamma_y) \gamma_5 d^s_c(\vec{x}, t)\right]$, where the superscript $s$ denotes smeared quark fields. The source position of $N$ is denoted by $(\vec{x}, t)$, $t_s$ is the source-sink separation, $\tau$ is the operator insertion time, and $\mathcal{P}^{\mathrm{3pt}}_{\alpha\beta}$ selects the nucleon spin polarization. For the helicity case, we use $\Gamma = i\gamma_5\gamma_z$ and $\mathcal{P}^{\mathrm{3pt}}_{\alpha\beta} = \frac{1}{2}(1+\gamma_t) i\gamma_5\gamma_z$; for the unpolarized case, we use $\Gamma = \gamma_t$ and $\mathcal{P}^{\mathrm{3pt}}_{\alpha\beta} = \frac{1}{2}(1+\gamma_t)$.

The lattice calculations are performed using 81 gauge-field configurations of the 2+1-flavor domain-wall fermion ensemble 64I~\cite{RBC:2023pvn}, with 5-dimensional volume $N_s^3 \times N_t \times L_5 = 64^3 \times 128 \times 12$, physical up, down, and strange quark masses, and spacing $a = 0.0836$~fm.

To improve ground-state overlap and reduce statistical noise, we employ boosted Gaussian smearing~\cite{Bali:2016lva} of the quark fields in the interpolation operator. Gaussian smearing is implemented with radius $r_G = 0.75$~fm and quark boost parameter $j_z = 0$ and 3~\cite{Gao:2021xsm, Gao:2020ito}, corresponding to nucleon momentum $P_z = 2\pi n_z / (N_s a)$ with $n_z = 0$ and 7, yielding $P_z^{\mathrm{max}} = 1.62$~GeV.

Quark propagators are computed from CG-fixed gauge configurations using a deflated solver with 2000 eigenvectors. For each gauge configuration, we perform multiple measurements of $C^{\mathrm{3pt}}_{\mathcal{P}, \Gamma}$ using the all-mode averaging technique~\cite{Shintani:2014vja}. For $n_z = 0$, we compute one exact solve~\cite{Aubin:2019usy,Blum:2016lnc} (solver tolerance $10^{-8}$) and 4 sloppy solves (solver tolerance $10^{-4}$) at $t_s/a = \{6, 8, 10\}$. For $n_z = 7$, we compute 1 exact and \{16, 32, 64, 128\} sloppy solves at $t_s/a = \{4, 6, 8, 10\}$, respectively.

We compute isovector ($u-d$) contributions for both helicity and unpolarized quasi-TMD beam functions. To study flavor dependence, the isoscalar ($u+d$) channel is also calculated for the unpolarized case, neglecting quark-line disconnected contributions. These contributions are expected to be small~\cite{Alexandrou:2021oih,Alexandrou:2020sml,Liang:2019xdx}, compared to the other systematic uncertainties. A dedicated investigation is needed in future work to quantify them.

\begin{figure}[th!]
    \centering
    \includegraphics[width=0.45\textwidth]{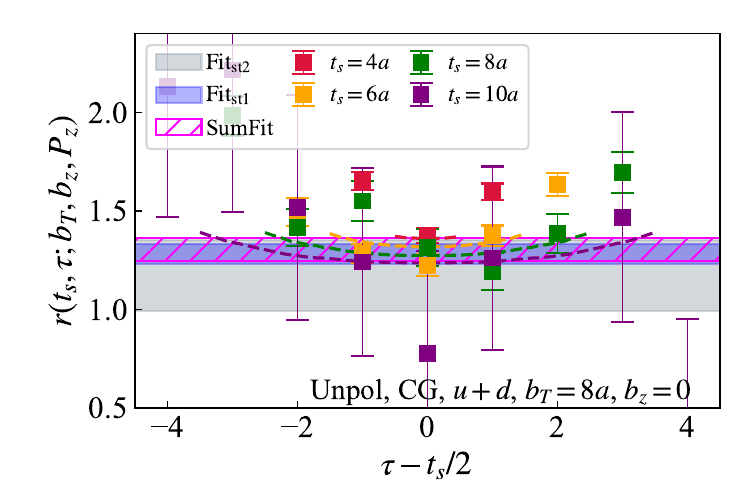}
    \includegraphics[width=0.45\textwidth]{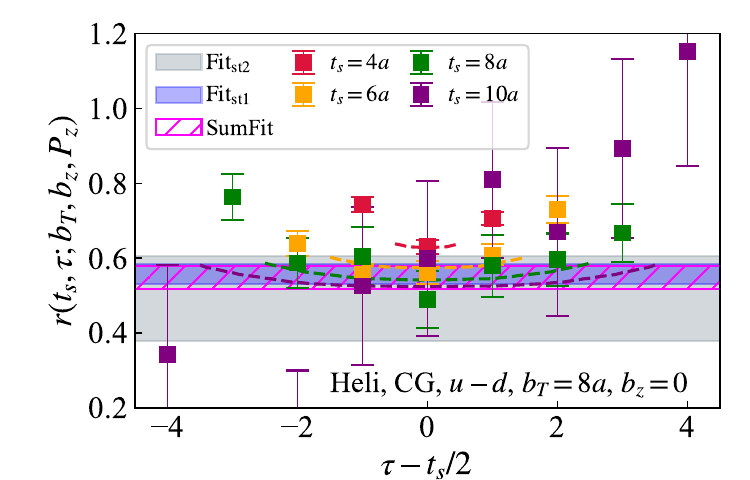}
    \caption{Ratios $r(t_s,\tau;b_T,b_z,P_z=1.62~\mathrm{GeV})$ for unpolarized (top) and helicity (bottom) CG quasi-TMD beam function matrix elements at $b_T = 8a$ and $b_z = 0$. Dashed lines show results from two-state fits; bands indicate the extracted values from $\mathrm{Fit}_{\mathrm{st1}}$, $\mathrm{Fit}_{\mathrm{st2}}$, and SumFit.\label{fig:ratiofit}}
\end{figure}

\emph{Results.}  We extract the bare matrix elements of the CG quasi-TMD beam functions by analyzing the ratio of three-point to two-point correlators,
\begin{align}\label{eq:3pt2ptratio}
\begin{split}
	r(t_s,\tau;b_T,b_z,P_z) &= \frac{C^{\mathrm{3pt}}(t_s,\tau;b_T,b_z,P_z)}{C^{\mathrm{2pt}}(t_s;P_z)},
\end{split}
\end{align}
where $C^{\mathrm{2pt}}$ denotes the two-point correlator of the unpolarized proton. In the limit $\tau, t_s-\tau \rightarrow \infty$, the excited-state contributions are exponentially suppressed, and the ratio isolates the ground-state matrix elements $\tilde{h}^B(b_T, b_z, P_z, a)$. For finite $t_s$ and $\tau$, we apply three extraction strategies: one-state/plateau fit ($\mathrm{Fit}_{\mathrm{st1}}$), two-state fit ($\mathrm{Fit}_{\mathrm{st2}}$), and summation fit (SumFit).

In $\mathrm{Fit}_{\mathrm{st1}}$, we fit a constant to the stable region of $r$ for $t_s = 6a, 8a, 10a$, excluding up to $\tau = 3a$ at both ends of each source-sink separation. For $\mathrm{Fit}_{\mathrm{st2}}$, we use $\tau \in [2a, t_s - 2a]$ for joint fit of $t_s\in [4a, 10a]$ and truncate the spectral sum in both $C^{\mathrm{2pt}}$ and $C^{\mathrm{3pt}}$ to include only the ground and first excited states. We extract the two energy levels and corresponding overlap amplitudes from fits to $C^{\mathrm{2pt}}$, which are then used in the spectral decomposition of $r$ to fit the matrix elements. For SumFit, we use $r$ at $t_s = 4a, 6a, 8a, 10a$, summing over $\tau \in [2a, t_s - 2a]$, and extract the matrix element from the slope of a linear fit in $t_s$. All fits are performed with an uncorrelated $\chi^2$, with correlations preserved through configuration-wise bootstrap. More details can be found in Appendix~\ref{app:CGz}.

Figure~\ref{fig:ratiofit} shows representative results at $P_z = 1.62~\mathrm{GeV}$, $b_T = 8a$, and $b_z = 0$, for the unpolarized and helicity matrix elements. The consistency across different methods indicates that excited-state effects are under control. For the final analysis, we average over all three methods on a bootstrap sample-by-sample basis. Additional details and examples are provided in Appendix~\ref{app:CGz}.

\begin{figure*}[th!]
    \centering
    \includegraphics[width=0.32\textwidth]{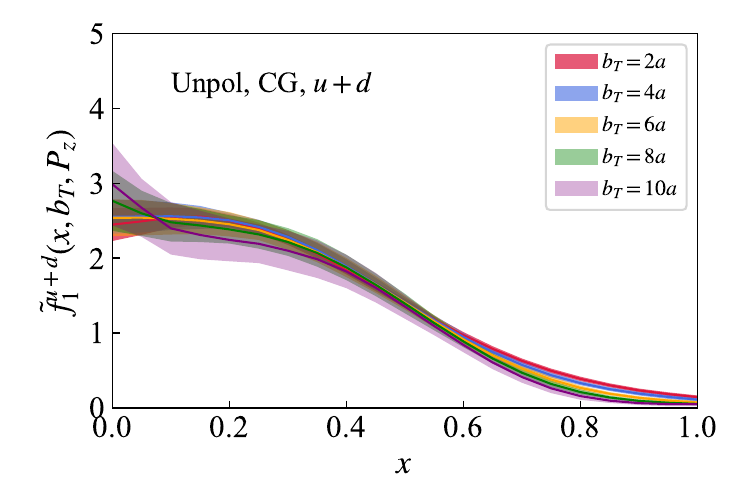}
    \includegraphics[width=0.32\textwidth]{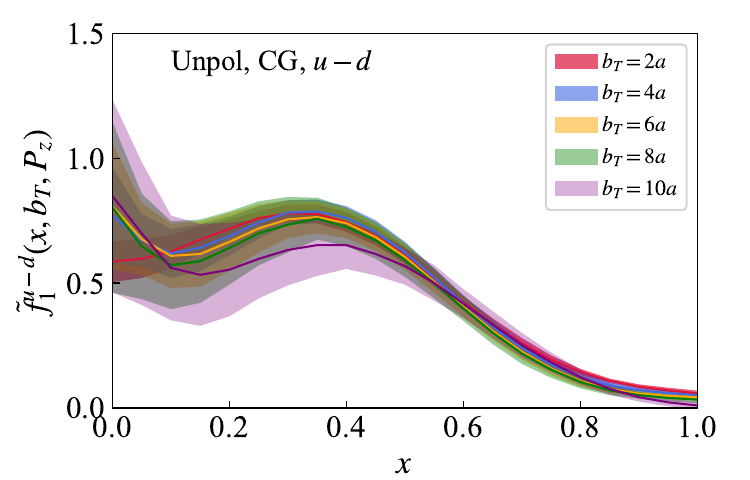}
    \includegraphics[width=0.32\textwidth]{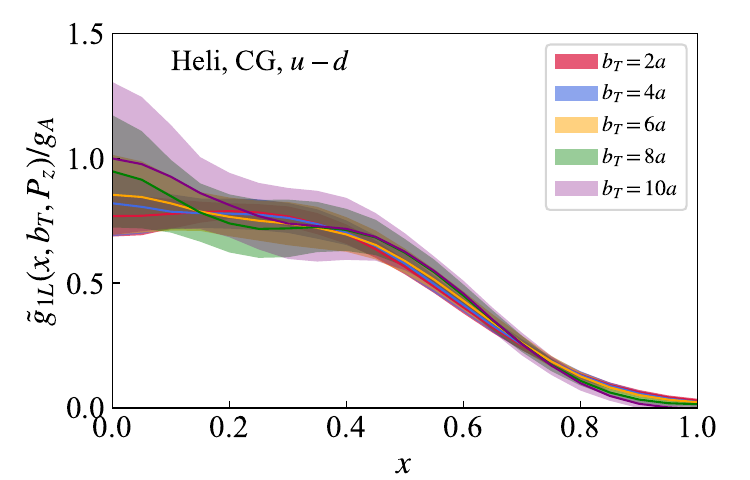}
    \caption{The $x$-dependent quasi-TMD beam functions are presented as functions of $x$ at several values of $b_T$. From left to right, the panels the isoscalar unpolarized, isovector unpolarized, and isovector helicity cases, respectively.
    \label{fig:quasiB}} 
\end{figure*}

The bare matrix elements require renormalization. For CG quasi-TMD beam functions, this reduces to a multiplicative constant independent of the quark separation $\{b_T, b_z\}$ and external hadron states~\cite{Gao:2023lny,Zhang:2024omt}. Following Refs.~\cite{Bollweg:2024zet,Avkhadiev:2024mgd}, we apply the ratio
\begin{align}\label{eq:rnm}
    \tilde{h}_i^R(b_T, b_z, P_z) = \mathcal{N} 
    \frac{\tilde{h}_i^{B}(b_T, b_z, P_z, a)}{\tilde{h}^{B,{u_v - d_v}}_{f_1}(b_T, 0, 0, a)},
\end{align}
where the numerator corresponds to any operator or flavor channel considered in this Letter, and the denominator is fixed to the isovector unpolarized matrix element to cancel the common ultraviolet divergence~\cite{Gao:2023lny}. $\mathcal{N} = \tilde{h}^{B,u_v-d_v}_{f_1}(0, 0, 0, a) / \tilde{h}_i^{B}(0, 0, P_z, a)$ is a normalization factor. Although this renormalization depends on $b_T$, it cancels in the ratios of quasi-TMD beam functions, which we present as the final results. Nevertheless, its inclusion is useful for reducing common systematic errors in intermediate steps such as discretization artifacts and excited-state contamination. This work focuses exclusively on the real part of the matrix element, which provides access to the valence-quark contribution $q_v$ in the unpolarized case and the combined quark–antiquark contribution $\Delta q_+$ in the helicity case. Previous studies~\cite{Gao:2023ktu,Gao:2022uhg,Fan:2020nzz} have shown that PDFs (or their even moments) extracted from the real part of the matrix elements agree well with global fits.

Due to confinement, the spatial correlators decay exponentially to zero at large $b_z$~\cite{Gao:2021dbh}, making the FT numerically stable. To ensure smooth behavior and suppress the noise at large $b_z$, we apply an exponential extrapolation for $b_z \gtrsim 1$~fm~\cite{Ji:2020brr}. The associated systematic uncertainty is estimated by varying the extrapolation range, which has negligible impact on the moderate-$x$ region. After renormalization and FT, we obtain the helicity quasi-TMD beam function $\tilde{g}_{1L}^{\Delta u_+ - \Delta d_+}(x, b_T, P_z)$ and the unpolarized quasi-TMD beam functions $\tilde{f}_1^{u_v - d_v}(x, b_T, P_z)$ and $\tilde{f}_1^{u_v + d_v}(x, b_T, P_z)$, shown in \fig{quasiB}. According to \Eq{qTMDfac}, the quasi-TMD beam function approximates the TMDPDF in $x$-space at $\zeta = (2xP_z)^2$, up to a $b_T$-dependent normalization factor---the soft function---as well as matching and power corrections. Since the renormalization factor in \Eq{rnm} introduces $b_T$ dependence, the renormalized quasi-TMD beam function is not monotonic in $b_T$. Nevertheless, we observe that it becomes increasingly suppressed as $b_T$ grows when $x \to 1$, indicating that the $O(1/(xP_z b_T))$ power correction is suppressed, which is consistent with previous lattice findings~\cite{LatticePartonLPC:2023pdv,Avkhadiev:2023poz,Avkhadiev:2023poz,Bollweg:2024zet,Bollweg:2025iol}. The statistical errors of the quasi-TMD beam function decrease with increasing $x$ and become small in the valence quark region. Although these errors grow with increasing $b_T$ due to the worsening signal-to-noise ratio, they remain sufficiently controlled to allow for meaningful predictions. Next, we construct the RGI ratios of the quasi-TMD beam functions. As a cross-check, we also compute the gauge-invariant (GI) quasi-TMD beam functions using the same lattice setup, and confirm that their RGI ratios are consistent with those from the CG method, as expected from \Eq{qTMDfac}~\cite{Ebert:2020gxr,Ebert:2022fmh,Zhao:2023ptv}; see Appendix~\ref{app:GICG}. 

\begin{figure}[h!]
    \centering
    \includegraphics[width=0.45\textwidth]{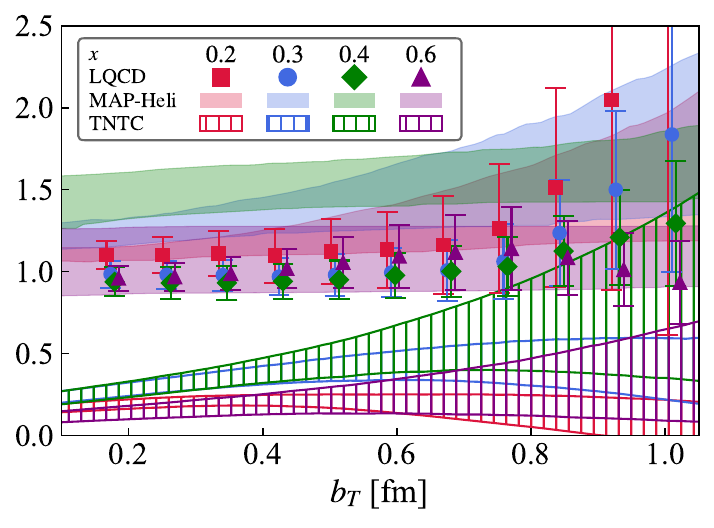}
	\caption{Lattice QCD results (filled symbols with error bars) for the ratio $R^{u-d}_{g_{1L}/f_1}(x, b_T)$ of isovector helicity to isovector unpolarized TMDPDFs in \Eq{HeliUnpol}, shown as a function of $b_T$ at several values of $x$. The hatched bands indicate the corresponding global fits from MAP-Heli~\cite{Bacchetta:2024yzl} and TNTC~\cite{Yang:2024drd}.\label{fig:HeliUnpol}}
\end{figure}
\begin{figure}[h!]
    \centering
    \includegraphics[width=0.45\textwidth]{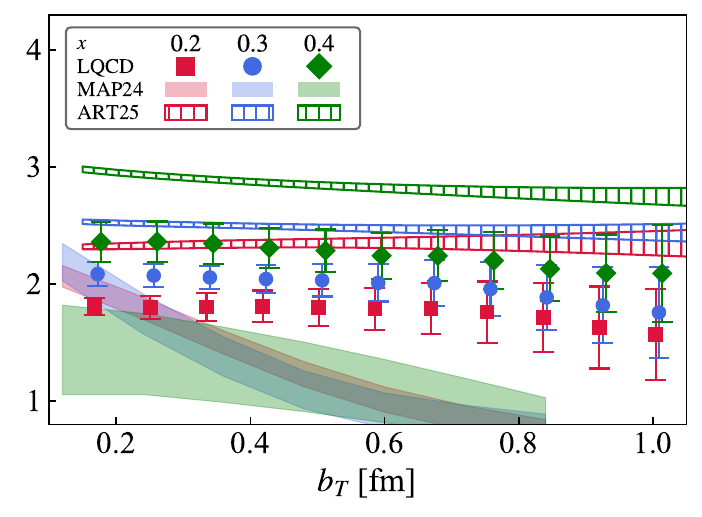}
	\caption{Lattice QCD results (filled symbols with error bars) for the ratio $R_{f_1}^{u/d}(x, b_T)$ of up- to down-quark unpolarized TMDPDFs in \Eq{UnpolUD}, shown as a function of $b_T$ at several values of $x$. The hatched bands denote the corresponding global fits from MAP24~\cite{Bacchetta:2024qre} and ART25~\cite{Moos:2025sal}.\label{fig:UnpolUD}}
\end{figure}

Figure~\ref{fig:HeliUnpol} shows the RGI ratio $R^{u-d}_{g_{1L}/f_1}(x, b_T)$, defined in Eq.~\eqref{eq:HeliUnpol}, for several values of $x$. Across all $x$ in the range $[0.2, 0.6]$, the ratio is consistent with being constant in $b_T$ within errors. This indicates that within the Collins-Soper-Sterman (CSS)~\cite{Collins:1984kg} formalism of parametrizing the TMDPDFs, the intrinsic nonperturbative TMD part is not sensitive to spin.
At smaller $x$, particularly around $x = 0.2$, our results are consistent with the MAP-Heli fit~\cite{Bacchetta:2024yzl}, while some deviations emerge at larger $x$. However, the TNTC fit~\cite{Yang:2024drd} yields substantially lower values across the entire $x$ range, although its $b_T$-dependence is also mild within uncertainties. The observed discrepancies between MAP-Heli and TNTC could be due to the experimental precision, analysis method and model uncertainties. While our lattice calculation is performed at a single lattice spacing and momentum, and thus subject to discretization effects and power corrections, these effects have been shown to be under control in similar LaMET studies, such as the pion valence quark PDF~\cite{Gao:2022iex}, the CS kernel~\cite{Bollweg:2024zet}, and the proton unpolarized PDF~\cite{Gao:2022uhg}. Therefore, our lattice results in the moderate-$x$ region can meaningfully differentiate and constrain phenomenological parametrizations of the TMDPDFs.

Figure~\ref{fig:UnpolUD} shows the ratio $R_{f_1}^{u/d}(x, b_T)$ of the valence $u$ to $d$ unpolarized TMDPDFs, as defined in \Eq{UnpolUD}, for several values of $x$. For comparison, results from the flavor-dependent global fits MAP24~\cite{Bacchetta:2024qre} and ART25~\cite{Moos:2025sal} are shown as bands. Better agreement in the trends in $b_T$ is observed between lattice and ART25, which also show a flat $b_T$-dependence across all $x$. Nevertheless, the magnitudes of the two results are noticeably different, which is subject to their own systematic uncertainties. At small $b_T$, some similarity between the lattice results and MAP24 emerges, likely due to the dominance of input collinear PDFs in their analysis. The mild $b_T$ variation in our lattice results indicates that the intrinsic nonperturbative TMD part of CSS parameterization has a weak dependence on the quark flavor.
Notably, while the $b_T$-dependence appears flat, the ratio still shows a noticeable dependence on $x$, indicating nontrivial longitudinal structure. Our finding is particularly valuable in light of the tensions among existing global fits, where incorporating flavor dependence introduces additional parameters that may increase statistical instability and contribute to systematic uncertainty. Therefore, our lattice QCD results offer a first-principles benchmark to guide future global analyses and help reduce the associated uncertainties.

\emph{Conclusion.} We presented the first lattice QCD calculation of the isovector helicity TMDPDF and the flavor-dependent unpolarized TMDPDFs for up and down quarks in the proton. This calculation used domain wall fermion configurations with physical quark masses, and the CG method was employed to reliably access the deep nonperturbative region up to $b_T\sim 1$ fm.

We obtain ratios of the TMDPDFs, which can be directly compared to those from global analyses. Our results show a close similarity between the helicity and unpolarized isovector TMDPDFs at moderate $x$, as well as a mild flavor dependence in the unpolarized case. 
With the current precision, our results can differentiate phenomenological parametrizations of the TMDPDFs and be used as valuable inputs to constrain future global analyses.

The present study provides first benchmarks at moderate~$x$ on a single lattice spacing and volume, and neglects disconnected diagrams in the unpolarized channel. Future work will extend these results by increasing the proton momentum, employing finer lattices, and addressing additional systematics—including excited-state contamination, disconnected diagrams, and finite-volume effects—so as to reach precision-level control. In parallel, together with a lattice determination of the soft function~\cite{Ji:2019sxk,Bollweg:2025iol}, these improvements will enable precise predictions of the proton’s three-dimensional structure with full spin dependence.

\emph{Acknowledgements.} This material is based upon work supported by the U.S.~Department of Energy, Office of Science, Office of Nuclear Physics through Contract No.~DE-SC0012704, Contract No.~DE-AC02-06CH11357, and within the frameworks of Scientific Discovery through Advanced Computing (SciDAC) award Fundamental Nuclear Physics at the Exascale and Beyond and the Topical Collaboration in Nuclear Theory Quark Gluon Tomography (QGT) with Award DE-SC0023646.

This research used awards of computer time provided by the INCITE program at Argonne Leadership Computing Facility, a DOE Office of Science User Facility operated under Contract DE-AC02-06CH11357; the ALCC program at the Oak Ridge Leadership Computing Facility, which is a DOE Office of Science User Facility supported under Contract DE-AC05-00OR22725; and the National Energy Research Scientific Computing Center, a DOE Office of Science User Facility supported by the Office of Science of the U.S.~Department of Energy under Contract DE-AC02-05CH11231, using NERSC award NP-ERCAP0028137 and NP-ERCAP0032114.

Our calculations were performed using the Grid~\cite{Boyle:2016lbp,Yamaguchi:2022feu} and GPT~\cite{GPT} software packages. We thank Christoph Lehner for his advice on using GPT. We thank the RBC and UKQCD collaborations for sharing the gauge field configurations used in our work.

%%%%%%%%%%%%%%%%%%%%%%%%%%%%%%%%%%%%%%%%%%%%%%%%%%%%%%%%%%%%%%%%%%%%%%
\appendix
\begin{widetext}
\section{Bare matrix elements of CG quasi-TMD beam functions in position-space}\label{app:CGz}

\begin{figure}[th!]
    \centering
    \includegraphics[width=0.45\textwidth]{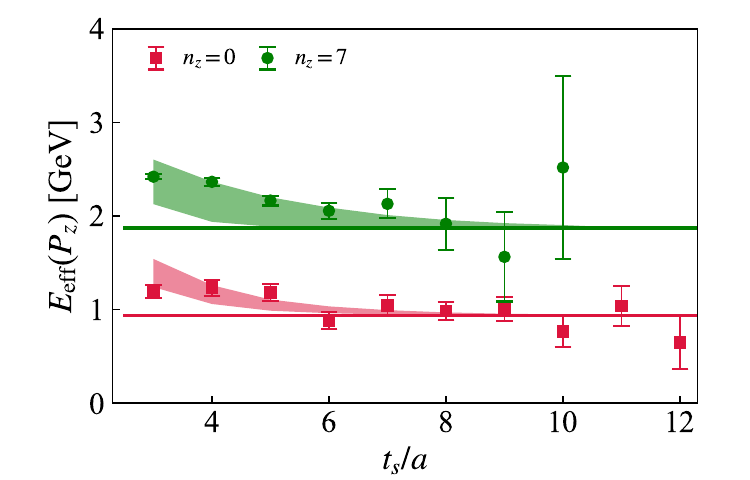}
	\caption{The effective mass obtained from two-point functions are shown. The lines are calculated from dispersion relation $E(P_z)=\sqrt{P_z^2+m_N^2}$ with $m_N=0.94$ GeV. The bands are constructed from the two-state fit results.\label{fig:Eeff}}
\end{figure}

\begin{figure}[th!]
    \centering
    \includegraphics[width=0.24\textwidth]{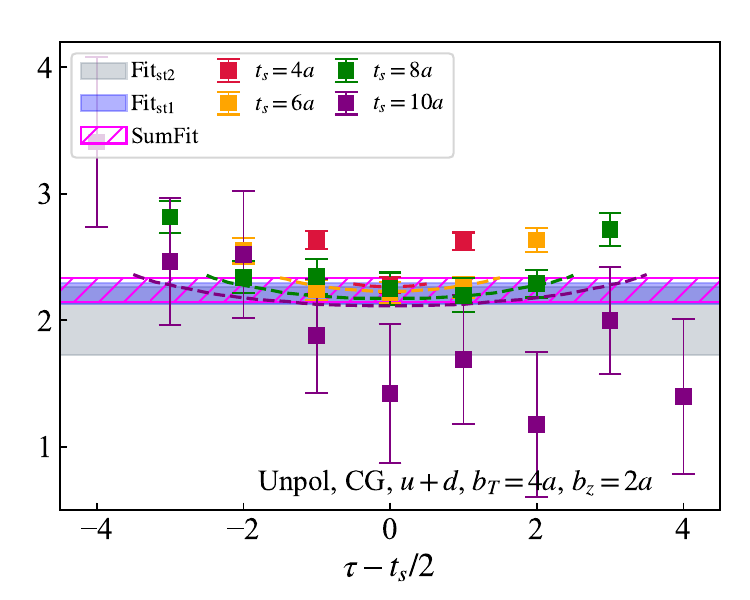}
    \includegraphics[width=0.24\textwidth]{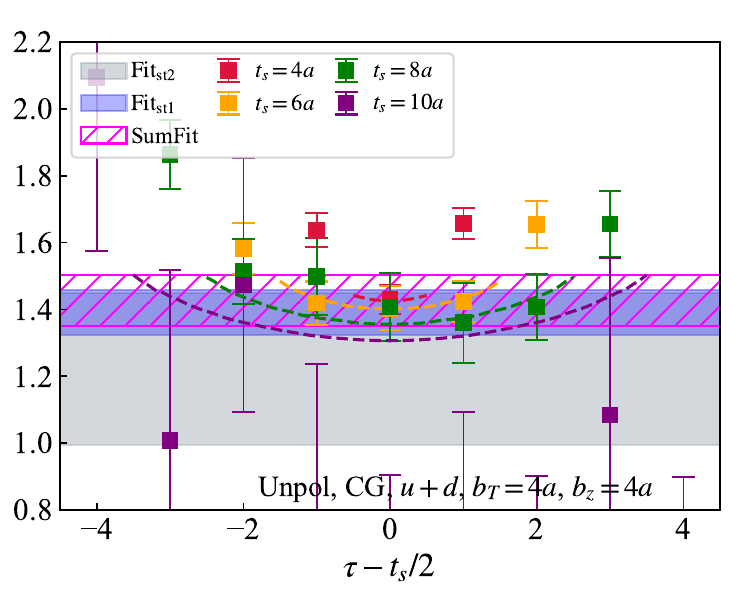}
    \includegraphics[width=0.24\textwidth]{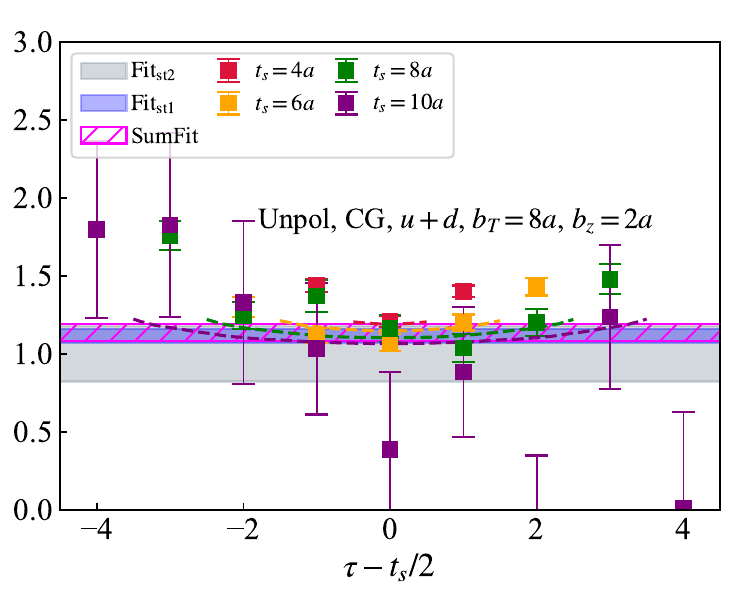}
    \includegraphics[width=0.24\textwidth]{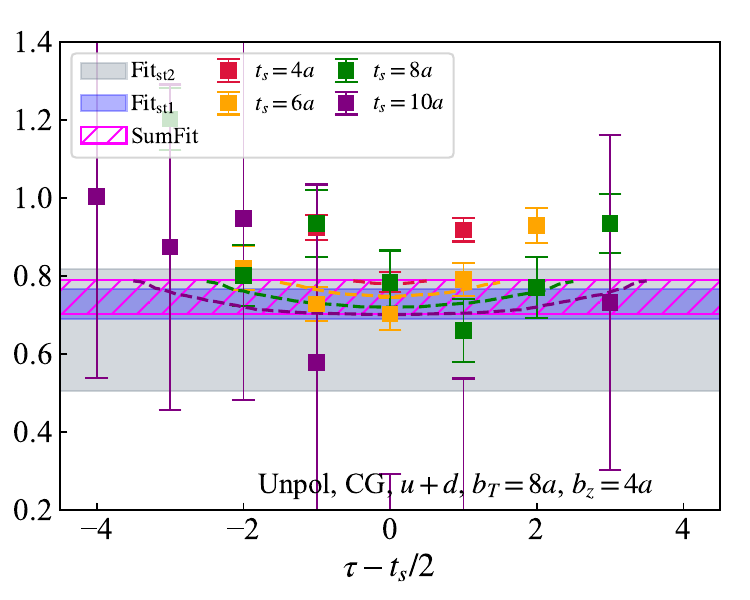}
    \includegraphics[width=0.24\textwidth]{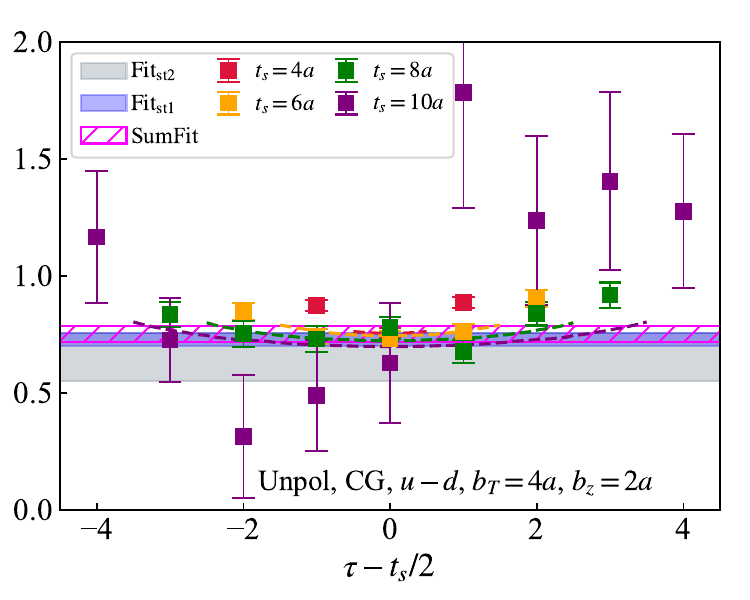}
    \includegraphics[width=0.24\textwidth]{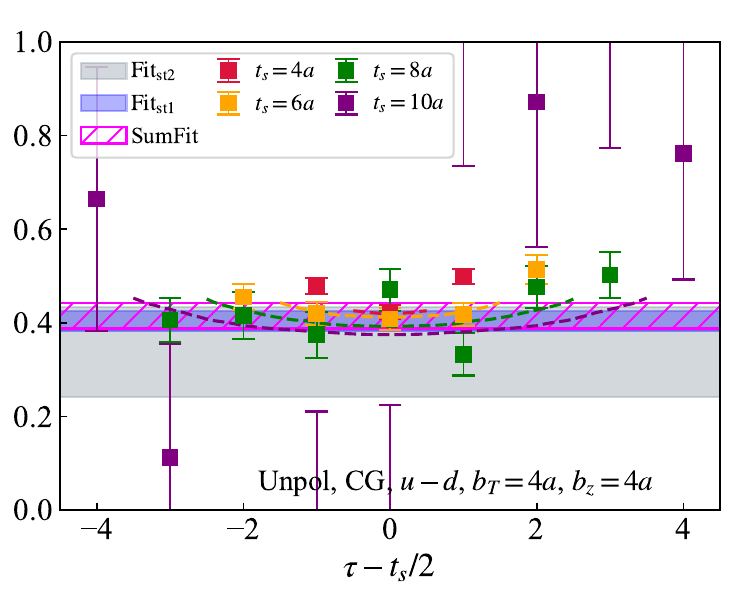}
    \includegraphics[width=0.24\textwidth]{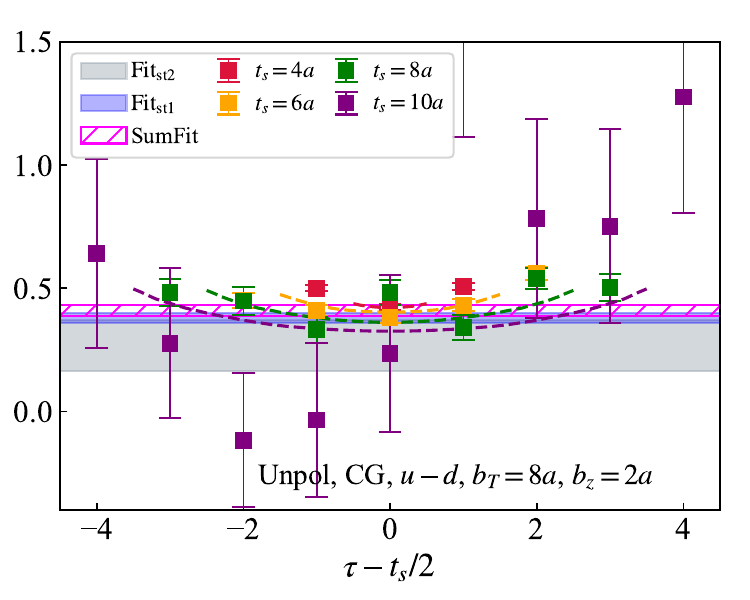}
    \includegraphics[width=0.24\textwidth]{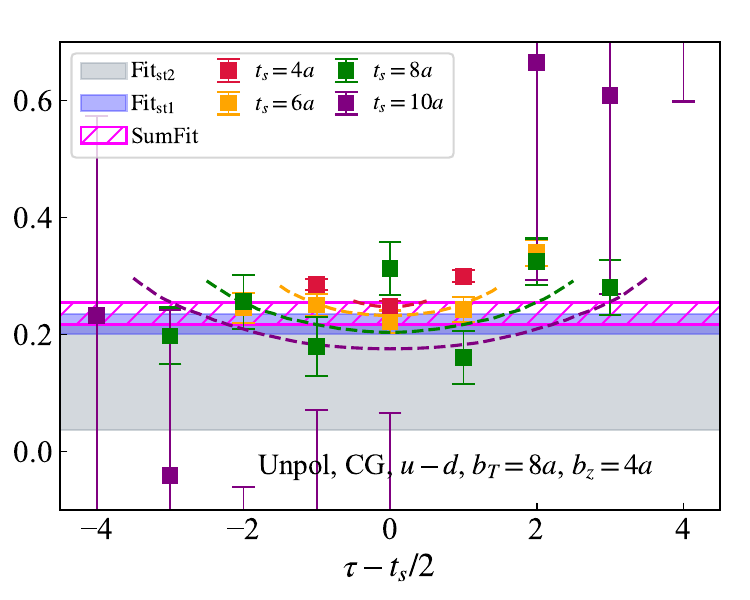}
    \includegraphics[width=0.24\textwidth]{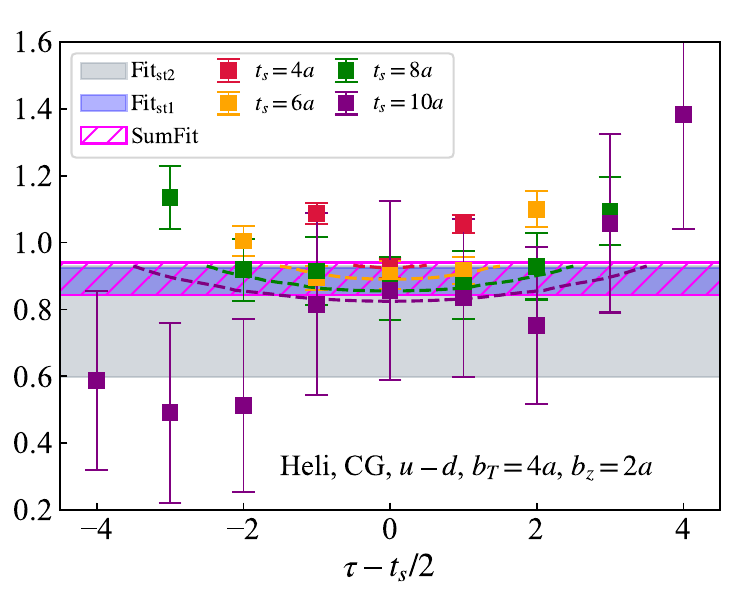}
    \includegraphics[width=0.24\textwidth]{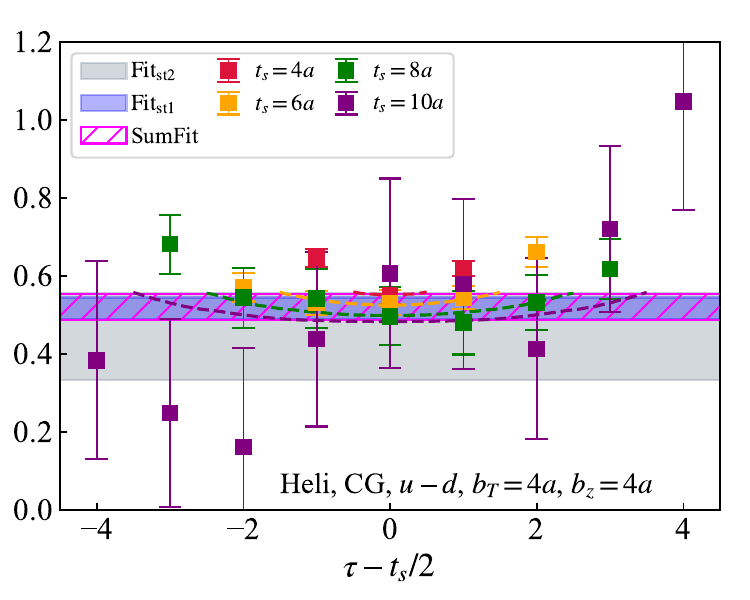}
    \includegraphics[width=0.24\textwidth]{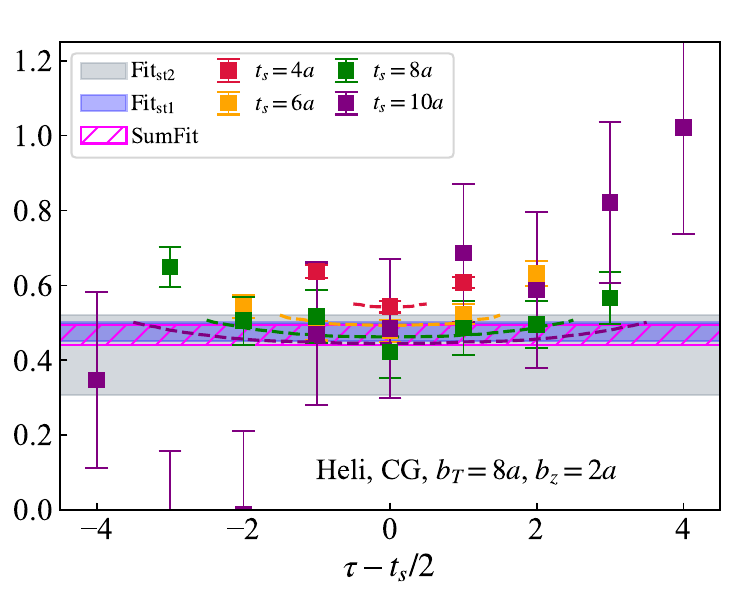}
    \includegraphics[width=0.24\textwidth]{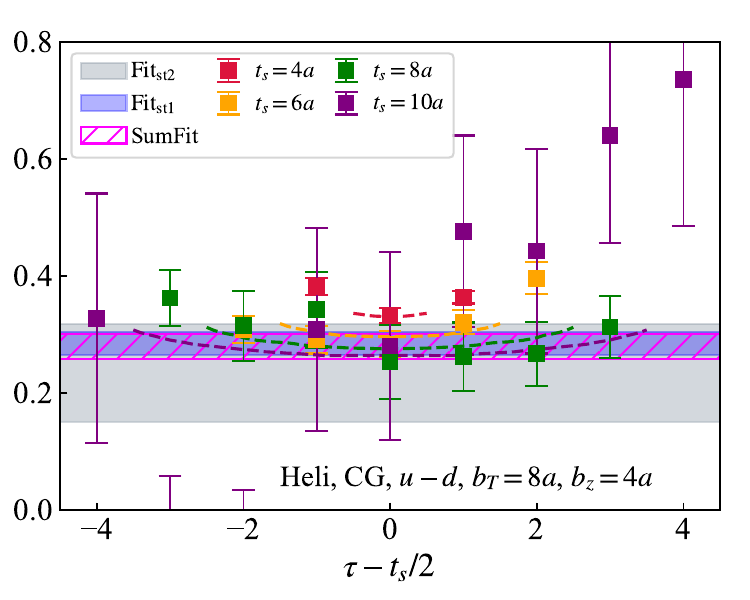}
    \caption{The ratios of three-point to two-point functions $r(t_s,\tau;b_T,b_z,P_z)$ for CG quasi-TMD beam functions with nucleon momentum $P_z=1.62$ GeV as functions of $t_s$ and $\tau$. From upper to lower panels, we show examples for iso-scalar unpolarized, iso-vector unpolarized, and iso-vector helicity cases respectively. The curves represent results from two-state fits, while the bands correspond to the results of one-state ($\rm Fit_{st1}$), two-state ($\rm Fit_{st2}$), and summation fits (SumFit), respectively.\label{fig:ratios}}
\end{figure}

\begin{figure}[th!]
    \centering
    \includegraphics[width=0.32\textwidth]{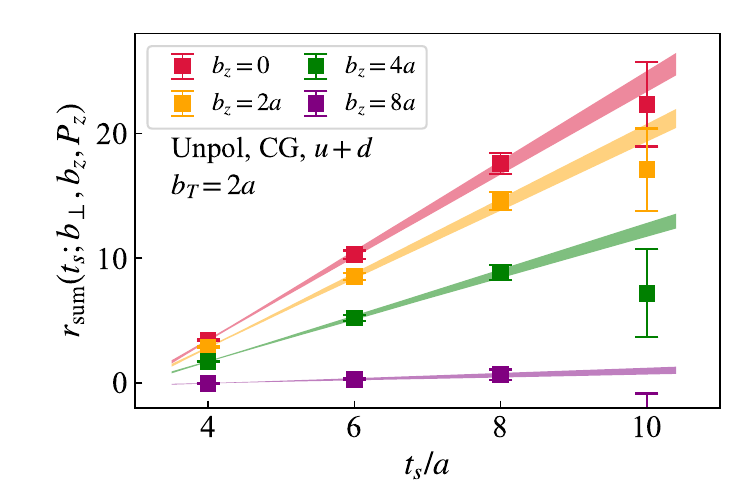}
    \includegraphics[width=0.32\textwidth]{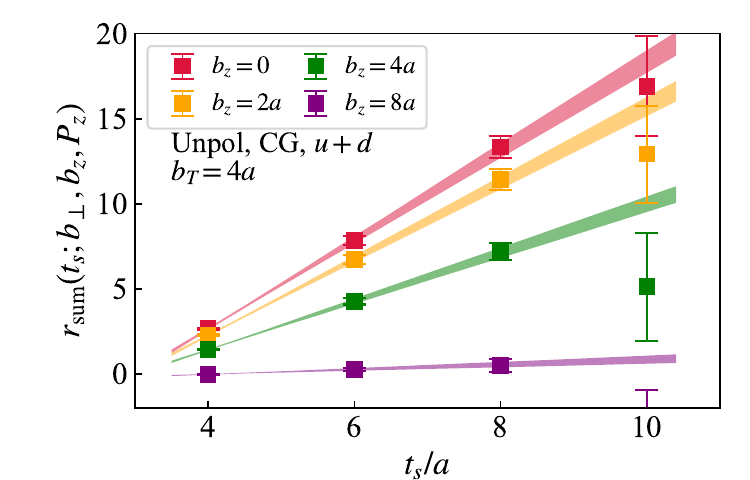}
    \includegraphics[width=0.32\textwidth]{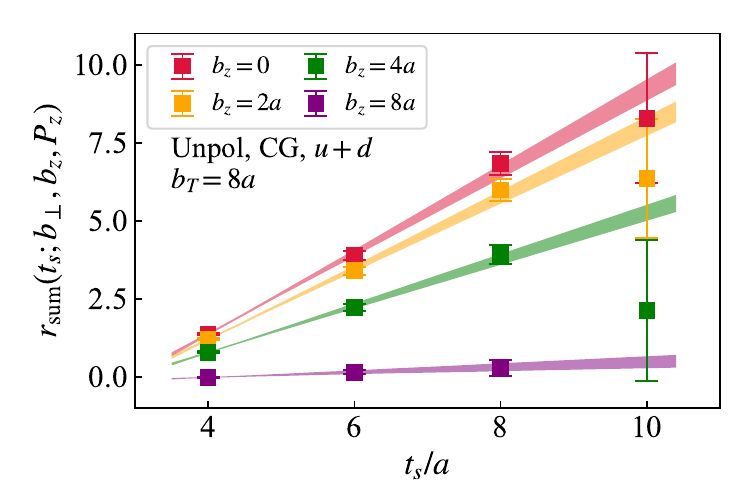}
    \includegraphics[width=0.32\textwidth]{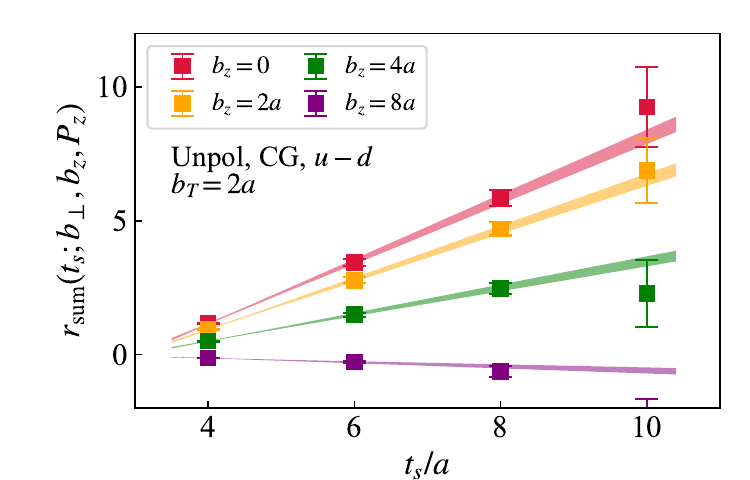}
    \includegraphics[width=0.32\textwidth]{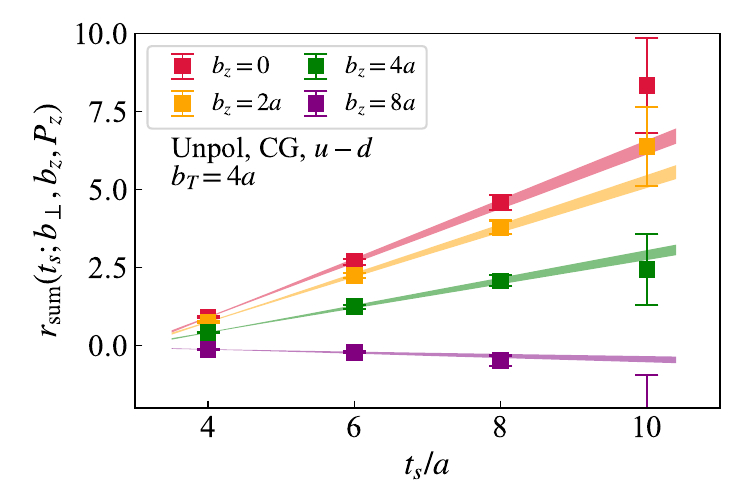}
    \includegraphics[width=0.32\textwidth]{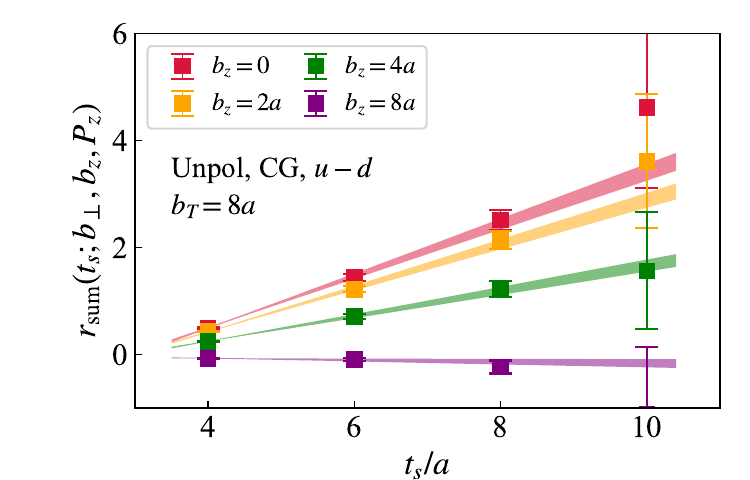}
    \includegraphics[width=0.32\textwidth]{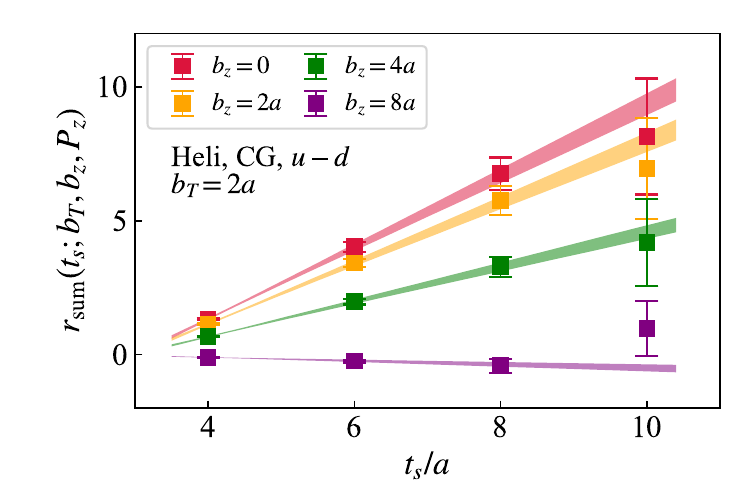}
    \includegraphics[width=0.32\textwidth]{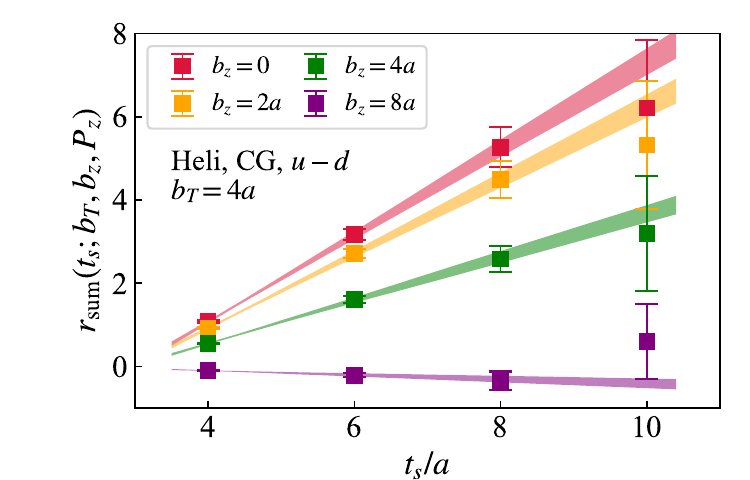}
    \includegraphics[width=0.32\textwidth]{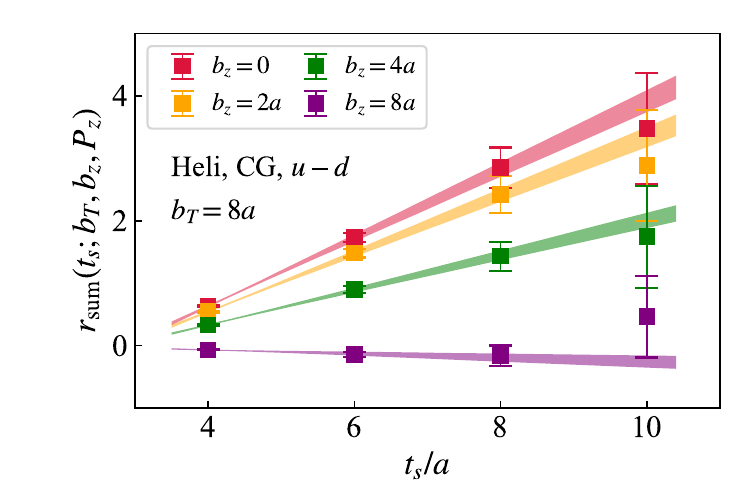}
    \caption{The summation fit results for CG quasi-TMD beam functions with nucleon momentum $P_z=1.62$ GeV as functions of $t_s$. From upper to lower panels, we show examples for iso-scalar unpolarized, iso-vector unpolarized, and iso-vector helicity cases respectively. From left to right, we present results of $b_T=2a,4a$ and $8a$ with multiple values of $b_z$. The bands are the fit results.\label{fig:sumfit}}
\end{figure}

\begin{figure}[th!]
    \centering
    \includegraphics[width=0.24\textwidth]{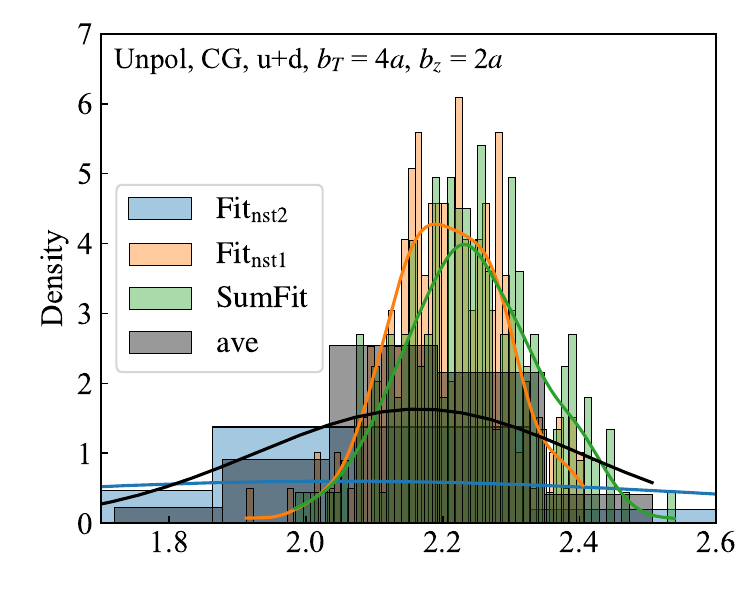}
    \includegraphics[width=0.24\textwidth]{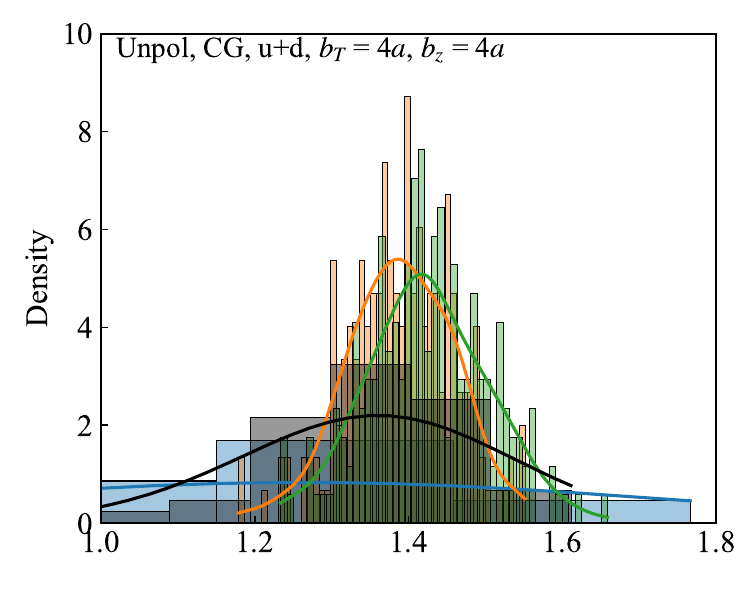}
    \includegraphics[width=0.24\textwidth]{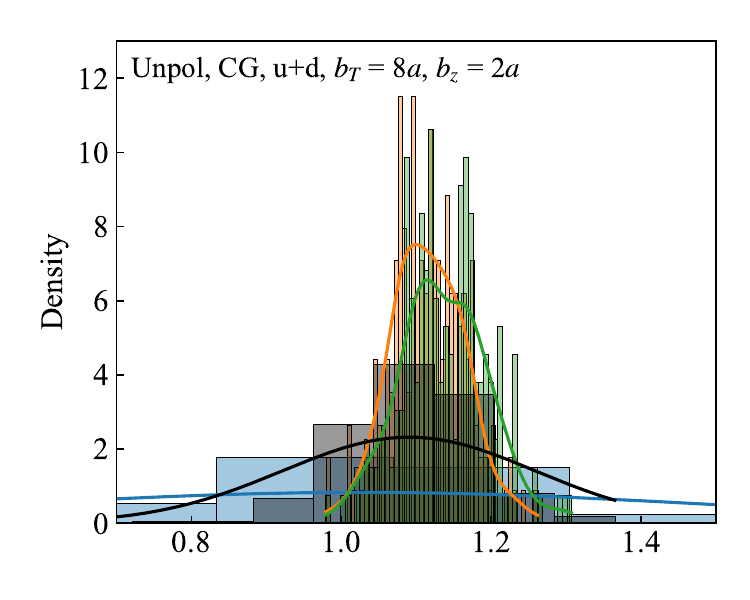}
    \includegraphics[width=0.24\textwidth]{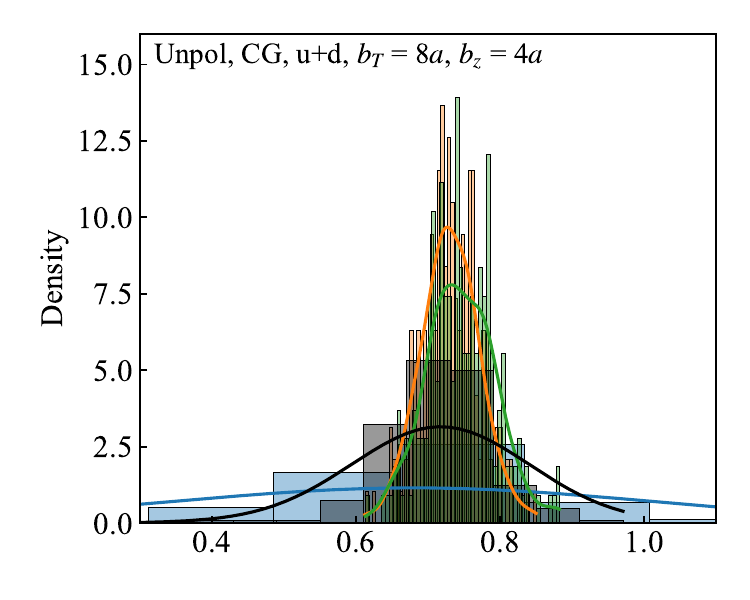}
    \includegraphics[width=0.24\textwidth]{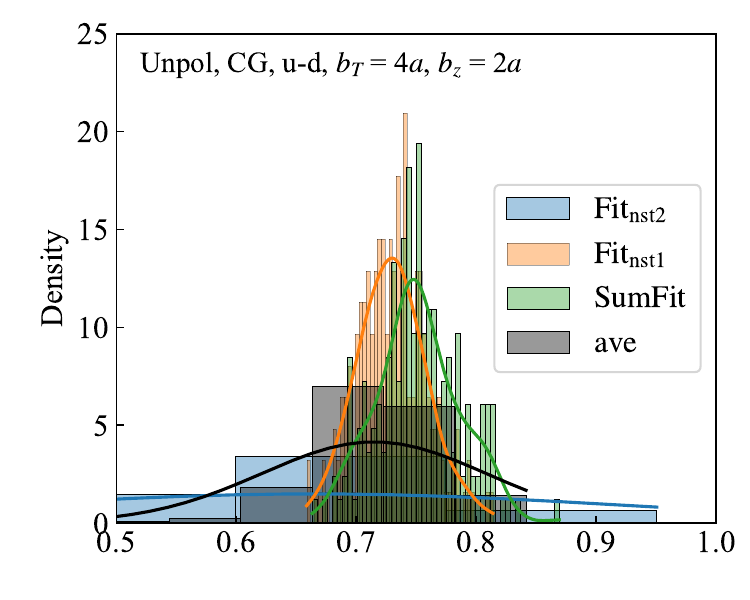}
    \includegraphics[width=0.24\textwidth]{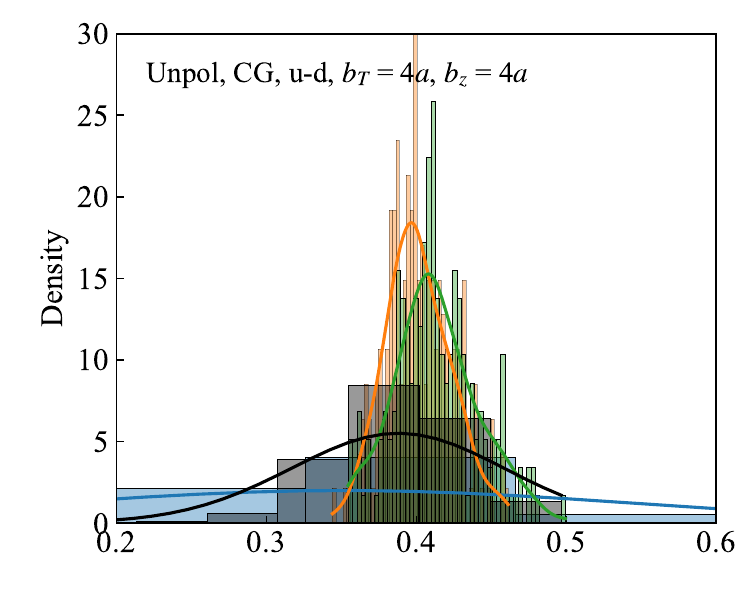}
    \includegraphics[width=0.24\textwidth]{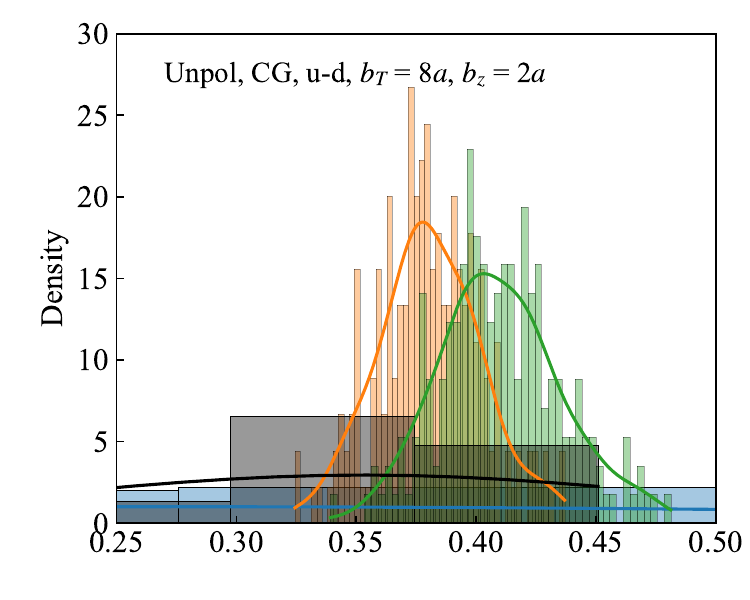}
    \includegraphics[width=0.24\textwidth]{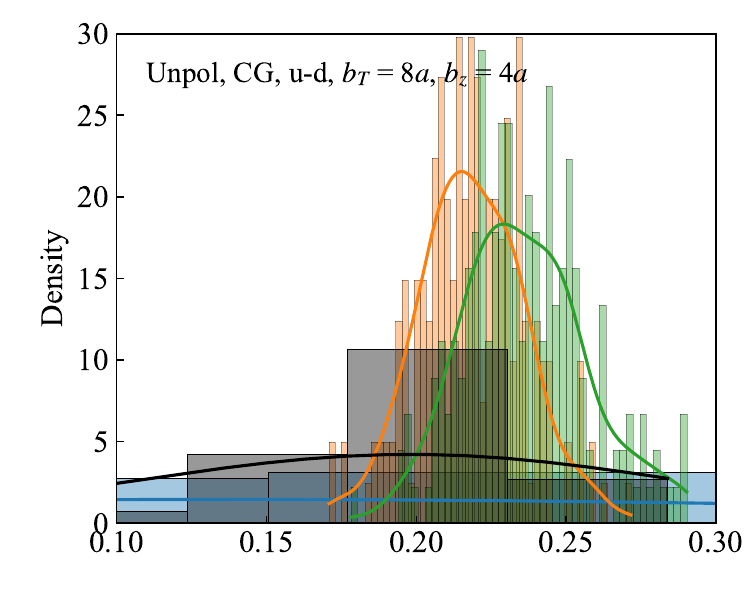}
    \includegraphics[width=0.24\textwidth]{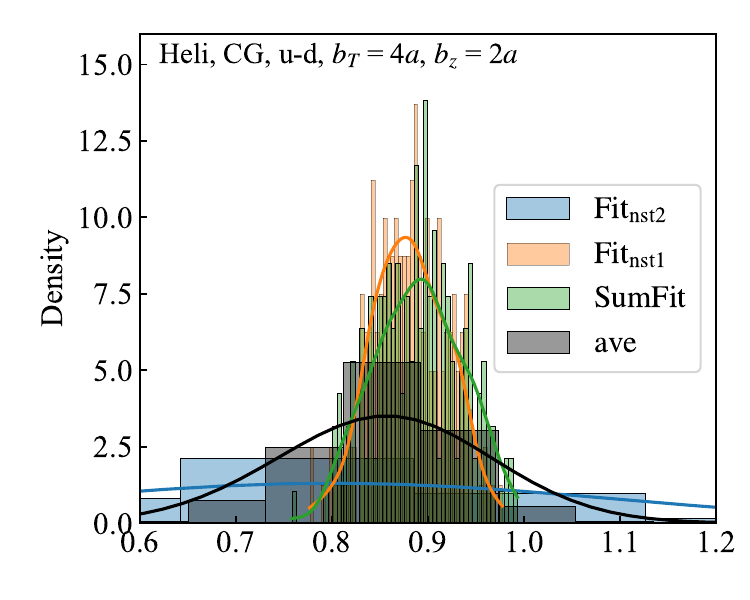}
    \includegraphics[width=0.24\textwidth]{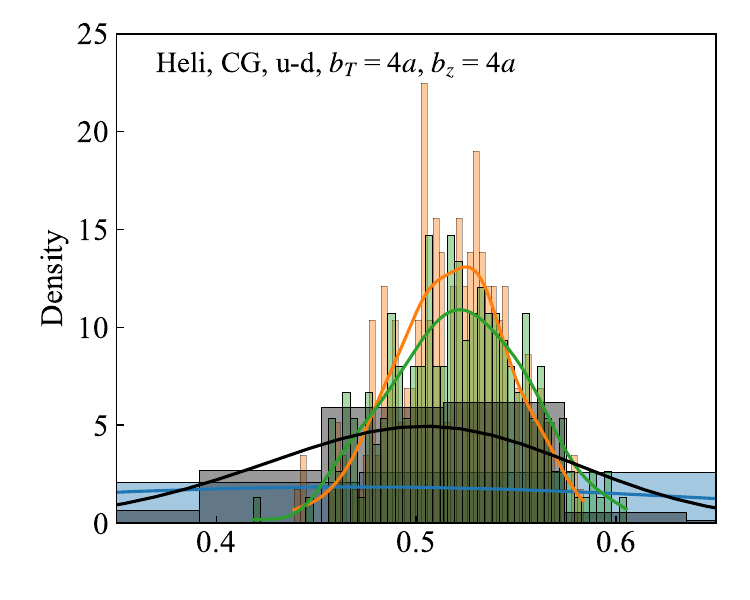}
    \includegraphics[width=0.24\textwidth]{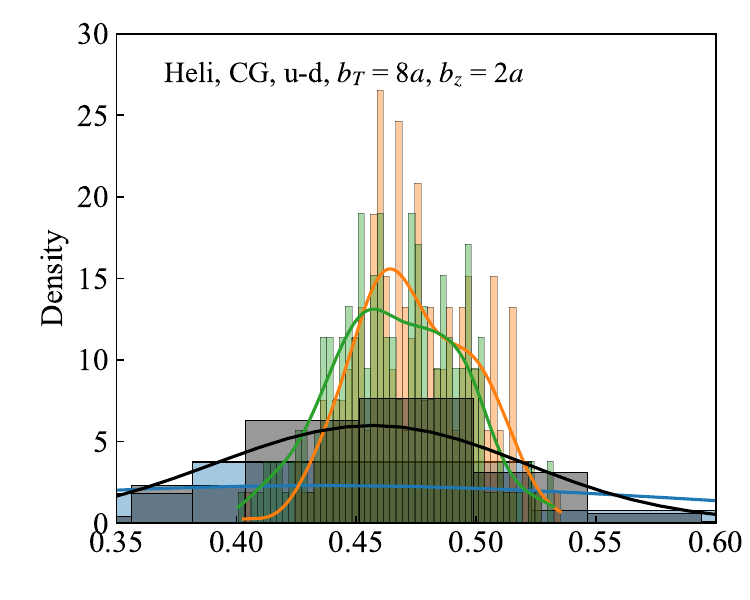}
    \includegraphics[width=0.24\textwidth]{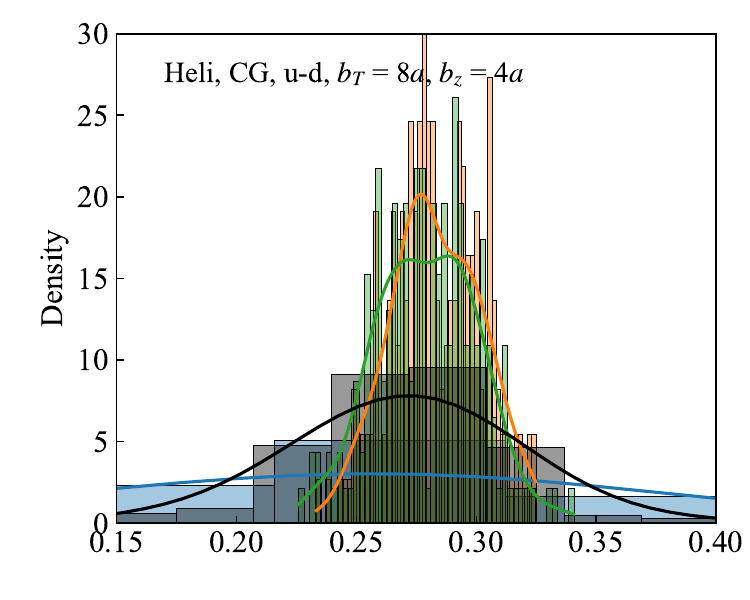}
    \caption{Bootstrap distributions of the bare matrix elements $\tilde{h}^B(b_T, b_z, P_z, a)$ of CG quasi-TMD beam functions at nucleon momentum $P_z=1.62~\mathrm{GeV}$, extracted using the SumFit, Fit$_\text{nst2}$, and Fit$_\text{nst1}$ methods together with their average (ave). From upper to lower panels, we show examples for iso-scalar unpolarized, iso-vector unpolarized, and iso-vector helicity cases respectively. The solid curves indicate kernel density estimates (KDE) of the underlying probability density.\label{fig:mxdistribution}}
\end{figure}

In this Appendix we present details of our analysis leading to the extractions of the bare matrix elements of CG quasi-TMD beam functions as function of $b_T$ and $b_z$.

As introduced in \Eq{3pt2ptratio} of the main text, the bare matrix elements can be extracted from the three-point to two-point correlation function ratio, $r(t_s, \tau; b_T, b_z, P_z)$. The three-point correlation function is defined in \Eq{3pt} of the main text. For the two-point correlation function, we compute,
\begin{align}\label{eq:2pt}
\begin{split}
    C^{\rm 2pt}_{\mathcal{P}} & (P_z, t_s) = \sum_{\vec{y}} e^{-i P_z \cdot (y_z - x_z)} \mathcal{P}^{\rm 2pt}_{\alpha \beta} \langle N^{(s)}_\alpha (\vec{y}, t_s + t) \overline{N}^{(s)}_\beta (\vec{x}, t)\rangle,
\end{split}
\end{align}
with $\mathcal{P}^{\rm 2pt} = \frac{1}{2} (1 + \gamma_t)$. Through the spectral decomposition, the two-point correlation functions can be expressed as,
\begin{align}\label{eq:c2ptsp}
    C^{\rm 2pt}(P_z, t_s) = \sum_{n=0}^{N_{\rm st}-1} \frac{|A_n|^2}{2 E_n} e^{-E_n t_s},
\end{align}
where $ E_n(P_z) $ represents the energy of the $ n $-th state, with $ n = 0, 1, \dots $ corresponding to the ground state, first excited state, and higher excited states. The overlap amplitude $ A_n = \langle n|\overline{N}^{(s)}(P_z)|{\Omega}\rangle $ quantifies the contribution of the nucleon interpolator to the $ n $-th state after boosted-momentum smearing. At large values of $t_s$, excited-state contributions are strongly suppressed, and the two-point function becomes dominated by the ground-state energy. In \fig{Eeff}, we show the effective mass, defined as,
\begin{align}
    E_{\rm eff}=-\log\left(\frac{C^{\rm 2pt}(P_z, t_s)}{C^{\rm 2pt}(P_z, t_s-1}\right)
\end{align}
evaluated from the two-point functions at $P_z = 0$ and $P_z = 1.62$~GeV as a function of $t_s$. As seen in the figure, for $t_s \gtrsim 6a$, the effective mass exhibits a plateau within statistical uncertainties. The lines in \fig{Eeff} correspond to the proton energy calculated from the dispersion relation, $E(P_z) = \sqrt{P_z^2 + m_N^2}$, with $m_N = 0.94$~GeV. These lines align well with the plateau region, indicating that the correlators are indeed dominated by the ground state, while excited-state contributions are suppressed and buried within the statistical noise.

The three-point correlation functions can also be expressed through the spectral decomposition,
\begin{align}\label{eq:c3ptsp}
\begin{aligned}
C^{\rm 3pt}_{\mathcal{P}, \Gamma}(t_s, \tau; b_T, b_z, P_z) = \sum_{m,n} A_m^* A_n \langle m|O|n \rangle e^{-\tau E_{n}} e^{-(t_s - \tau) E_m},
\end{aligned}
\end{align}
where $A_n$ and $E_n$ represent the same overlap amplitudes and energy levels as those in the two-point correlation functions. In the limit $t_s > \tau \rightarrow \infty$, the three-point to two-point correlation function ratio, $r(t_s, \tau; b_T, b_z, P_z)$, is dominiated by the ground-state matrix elements, yielding $\tilde{h}^B(b_T, b_z, P_z, a) = \langle 0 | O | 0 \rangle$.

\begin{figure}[th!]
    \centering
    \includegraphics[width=0.32\textwidth]{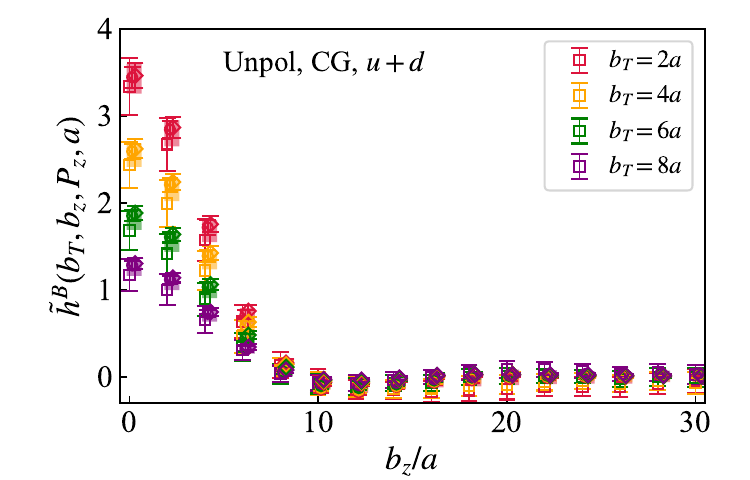}
    \includegraphics[width=0.32\textwidth]{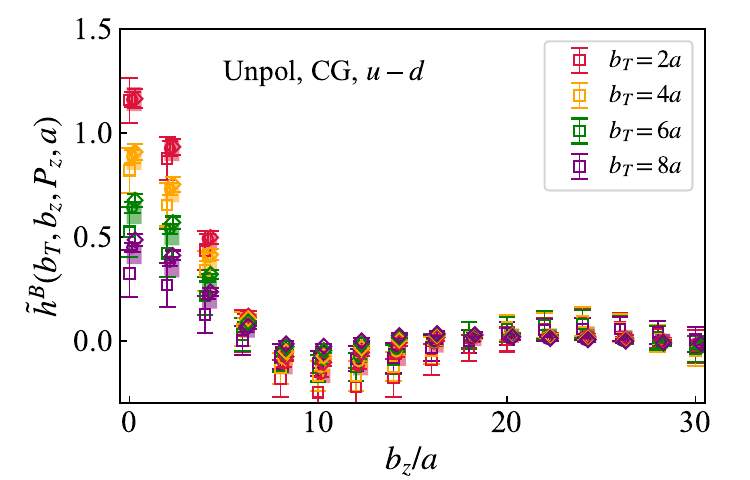}
    \includegraphics[width=0.32\textwidth]{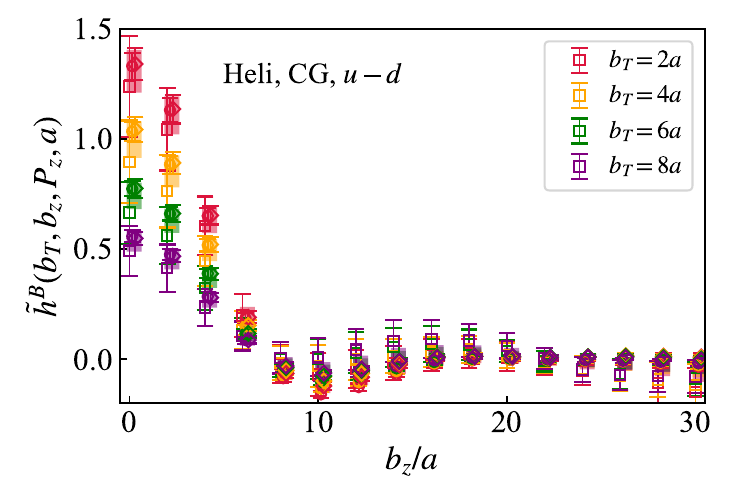}
    \caption{The real-part bare matrix elements for CG quasi-TMD beam functions with nucleon momentum $P_z=1.62$ GeV are shown as functions of $b_z$. The squared, circled and rhombus symbols represent results from two-state, one-state and summation fit respectively. The horizontal error bands are the averaged results obtained from all three methods. From left to right, we show examples for iso-scalar unpolarized, iso-vector unpolarized, and iso-vector helicity cases respectively.\label{fig:bmCG}}
\end{figure}

\begin{figure}[th!]
    \centering
    \includegraphics[width=0.4\textwidth]{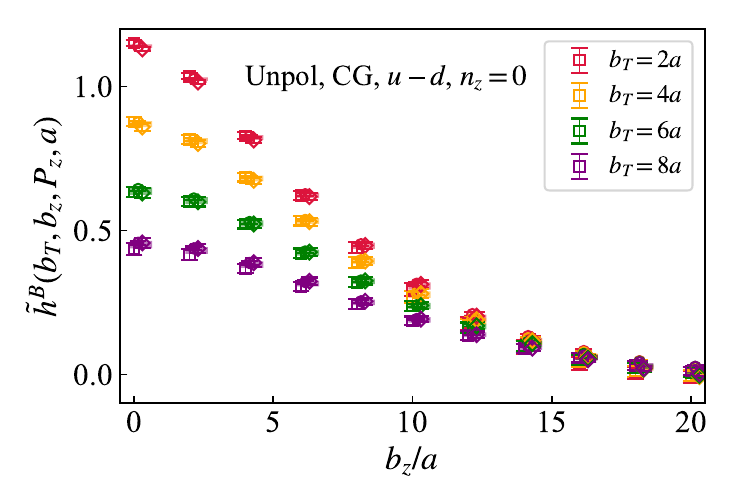}
    \caption{The bare matrix elements for CG iso-vector unpolarized quasi-TMD beam functions with nucleon momentum $P_z=0$ are shown as functions of $b_z$. The squared, circled and rhombus symbols represent results from two-state, one-state and summation fit respectively. The horizontal error bands are the averaged results obtained from all three methods.\label{fig:bmPz0}}
\end{figure}

In \fig{ratios}, we show representative ratio results for the case of $P_z = 1.62$ GeV with source-sink separations $t_s/a = 4, 6, 8, 10$. As seen in the plots, for larger $t_s$ and when the operator insertion time $\tau$ is near the midpoint $t_s/2$, the ratios exhibit minimal dependence on both $t_s$ and $\tau$. This behavior indicates that excited-state contamination is relatively mild in this region compared to the statistical uncertainties.

We first extract the bare matrix elements using a one-state (plateau) fit ($\mathrm{Fit}_{\mathrm{st1}}$), where a constant is fitted to the plateau region of the ratio. In this analysis, we omit the $t_s = 4a$ data and exclude three time insertions $\tau$ near each end of the source-sink separation to further suppress excited-state effects. The resulting fits are shown as blue bands in \fig{ratios}, and they are largely consistent with the data points included in the fit.

Next, we apply the summation method (SumFit), in which the ratio is summed over $\tau$ to further suppress excited-state contamination. The summed ratio is defined as:
\begin{align}
r_{\rm sum}(t_s; b_T, b_z, P_z) = \sum_{\tau = n_{\rm sk}a}^{t_s - n_{\rm sk}a} r(t_s, \tau; b_T, b_z, P_z),
\label{eq:ratioSum}
\end{align}
where its dependence on $t_s$ is modeled as:
\begin{align}
r_{\textup{sum}}(t_s) = (t_s - 2 n_{\textup{sk}} a) \tilde{h}^B(b_T, b_z, P_z, a) + B_0 + \mathcal{O}(e^{-(E_1 - E_0)t_s}).
\label{eq:sumModel}
\end{align}
In this analysis, we skip two time insertions ($n_{\rm sk} = 2$) at both ends of each $t_s$ for added suppression of excited-state effects. \fig{sumfit} shows the summation fit results for CG quasi-TMD beam functions at $P_z = 1.62$ GeV. From top to bottom, the panels present the iso-scalar unpolarized, iso-vector unpolarized, and iso-vector helicity cases, respectively. The bands represent straight lines reconstructed using \Eq{sumModel}, and the data points exhibit clear linear behavior, in excellent agreement with the fits. This confirms that $r_{\textup{sum}}(t_s)$ is largely governed by the ground-state matrix elements within statistical uncertainties.

We also perform a two-state fit ($\mathrm{Fit}_{\mathrm{st2}}$), which incorporates one excited state by truncating the spectral sums in \Eq{c2ptsp} and \Eq{c3ptsp} at $N_{\rm st} = 2$. We first determine the overlap amplitudes $A_n$ and energy levels $E_n$ from a two-state fit to the two-point functions. To stabilize the fit, we fix the ground-state energy $E_0$ to the value given by the dispersion relation and fit for $A_0$, $A_1$, and $E_1$. The effective masses reconstructed from this two-state fit are shown as bands in \fig{Eeff}, which successfully reproduce the $t_s$ dependence of the original data, validating the fit’s ability to capture excited-state behavior at shorter time separations. These amplitudes and energies are then used as priors in the two-state fit of the three-point function to two-point function ratios. Here, we also skip two time insertions ($n_{\rm sk} = 2$) at both ends of each $t_s$ to reduce excited-state contamination. In \fig{ratios}, the dashed curves represent results from the two-state fits and are in good agreement with the data.

Finally, we compare the bare matrix elements extracted from the one-state fit ($\mathrm{Fit}_{\mathrm{st1}}$), two-state fit ($\mathrm{Fit}_{\mathrm{st2}}$), and summation method (SumFit), shown as blue, grey, and purple bands in \fig{ratios}, respectively. The three fit strategies we employ---$\mathrm{Fit}_{\mathrm{st1}}$, $\mathrm{Fit}_{\mathrm{st2}}$, and SumFit---are complementary: $\mathrm{Fit}_{\mathrm{st1}}$ assumes negligible excited-state contributions and may be overly optimistic; $\mathrm{Fit}_{\mathrm{st2}}$ includes the first excited state but can become unstable and noisy with limited data precision; and SumFit is generally more robust, though it may still retain residual contamination at small $t_s$. To illustrate their statistical behavior, \fig{mxdistribution} shows the bootstrap distributions of the extracted matrix elements from all three methods, together with their sample-by-sample average. The substantial overlap of the distributions demonstrates consistency across fit strategies, while the narrower shape of SumFit and the broader tails of $\mathrm{Fit}_{\mathrm{st2}}$ reflect their differing sensitivities. In this Letter, we choose to average the results from all three methods, so that the spread among strategies is transparently propagated into the quoted uncertainties. Importantly, the results obtained from each method individually lie within the final error band, supporting the conclusion that the extracted matrix elements are dominated by the ground state, albeit with sizeable statistical uncertainties.

In \fig{bmCG}, we summarize the bare matrix elements extracted from the three methods for the CG case. The squared, circled and rhombus symbols represent results from two-state, one-state and summation fit respectively. Once again, good agreement across the three fitting methods is observed. The horizontal error bands are the averaged results obtained from all three methods.

For the renormalization in \Eq{rnm}, we require the zero-momentum bare matrix elements of iso-vector unpolarized quasi-TMD beam functions. These results are shown in \fig{bmPz0}.

\section{CG quasi-TMD beam functions in momentum-space} \label{app:CGx}

In this Appendix we present details of our analysis leading to the extractions of CG quasi-TMD beam functions as functions of $x$.

At large distances, spatial correlators are known to decay exponentially~\cite{Gao:2021dbh}. As shown in the upper panel of \fig{CG_qPDFx}, the renormalized matrix elements approach zero for $ b_z \gtrsim 1 $ fm, within the uncertainties. To perform the FT over $ b_z $ and extract the quasi-TMDs in $ x $-space, we model the long-distance behavior by fitting the data in the range $ b_z \in [b_z^L, b_z^L + 4a] $ using an exponential function, $A e^{-m b_z}$, with a prior constraint of $ m > 0.1 $ GeV. In the upper panel of \fig{CG_qPDFx}, we illustrate the fit results for $b_z^L=16a$ as the bands, which effectively capture the data behavior, although the uncertainties decrease more rapidly than the data itself. To account for potential systematic uncertainties arising from the extrapolation, we vary the fitting range by adjusting $b_z^L$ between $10a$ and $16a$. Finally, to compute $\tilde{f}(x, b_T, P_z)$, we combine a discretized FT for the range $b_z \in [0, b_z^L]$ with an analytical FT for the extrapolated region $b_z \in [b_z^L, \infty]$. This hybrid approach ensures a smooth and reliable reconstruction of the quasi-TMDs in momentum space.

In the lower panel of \fig{CG_qPDFx}, we compare results obtained with different choices of $b_z^L$, as well as with a naive DFT performed without extrapolation. The results from different $b_z^L$ values show excellent agreement, and the differences between the extrapolated and non-extrapolated (DFT-only) cases are minor, although the extrapolated results appear somewhat smoother. This mild difference arises because the matrix elements have already decayed to near-zero beyond $b_z \gtrsim 1$ fm. As discussed in the main text, this ensures that the effect of extrapolation and the associated model dependence is minimal.

\begin{figure}[th!]
    \centering
    \includegraphics[width=0.32\textwidth]{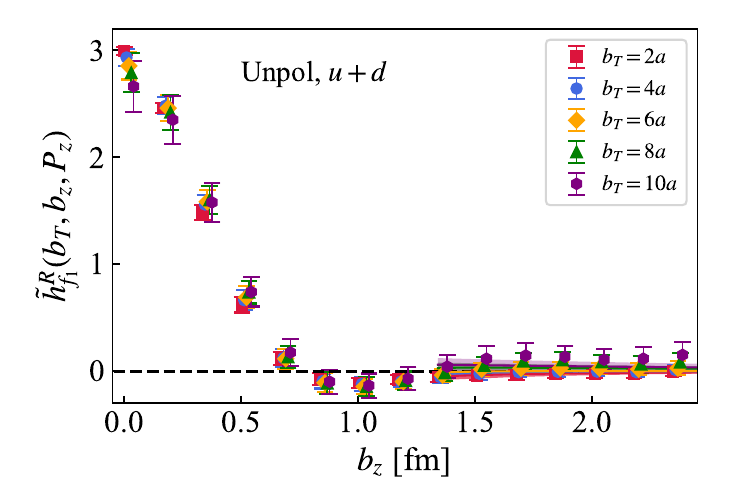}
    \includegraphics[width=0.32\textwidth]{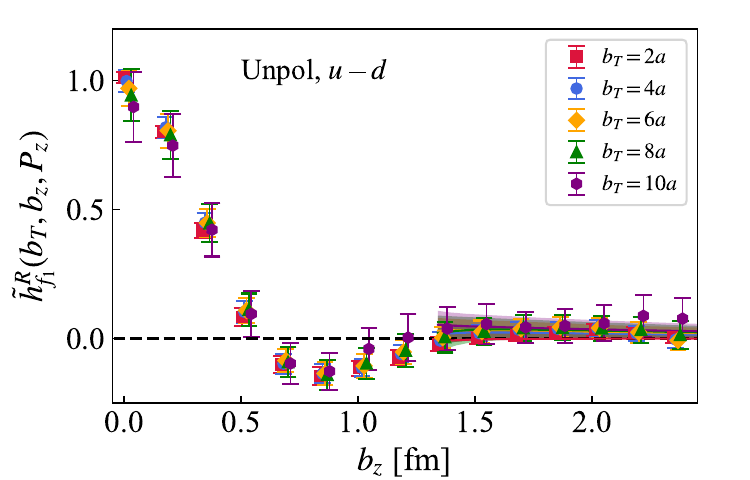}
    \includegraphics[width=0.32\textwidth]{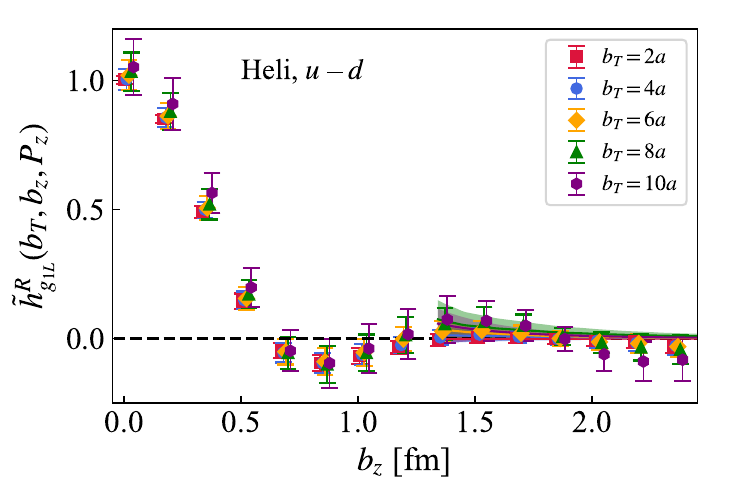}
    \includegraphics[width=0.32\textwidth]{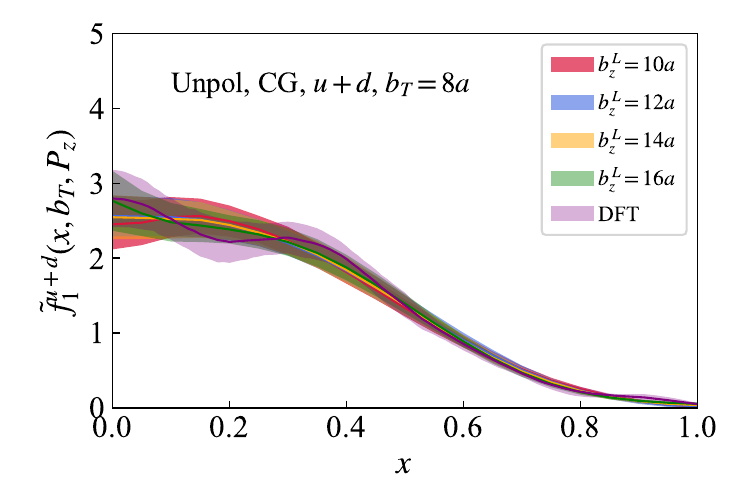}
    \includegraphics[width=0.32\textwidth]{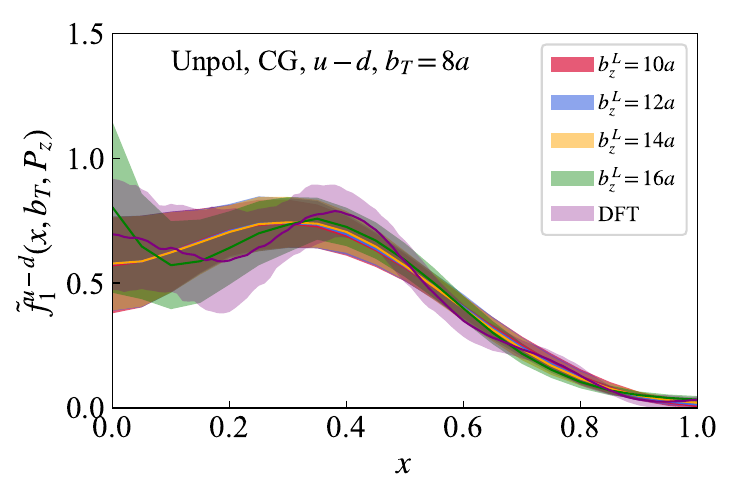}
    \includegraphics[width=0.32\textwidth]{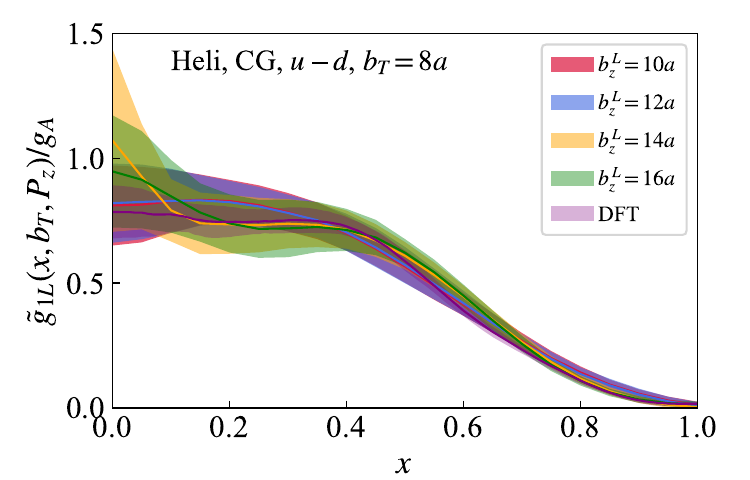}
    \caption{Upper panel: Real parts of the renormalized matrix elements for Coulomb-gauge (CG) quasi-TMD beam functions at nucleon momentum $P_z = 1.62$~GeV, plotted as functions of $b_z$ for several values of $b_T$. The bands indicate exponential fits used for extrapolation beyond $b_z^L = 16a$. Lower panel: $x$-dependent quasi-TMD beam functions at fixed $b_T = 8a$, obtained via discrete FT (DFT) with and without exponential extrapolation. Results are shown for $b_z^L$ ranging from $10a$ to $16a$. From left to right, examples are shown for the iso-scalar unpolarized, iso-vector unpolarized, and iso-vector helicity cases, respectively.\label{fig:CG_qPDFx}}
\end{figure}

\section{Gauge-invariant quasi-TMDPDFs}\label{app:GICG}
\begin{figure}[th!]
    \centering
    \includegraphics[width=0.32\textwidth]{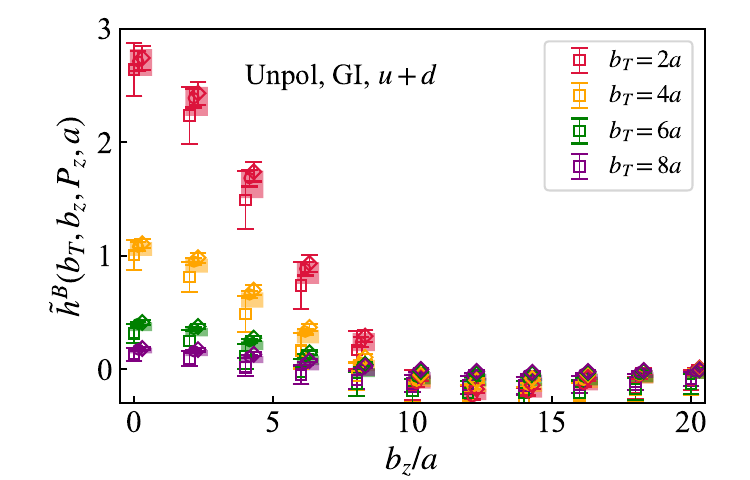}
    \includegraphics[width=0.32\textwidth]{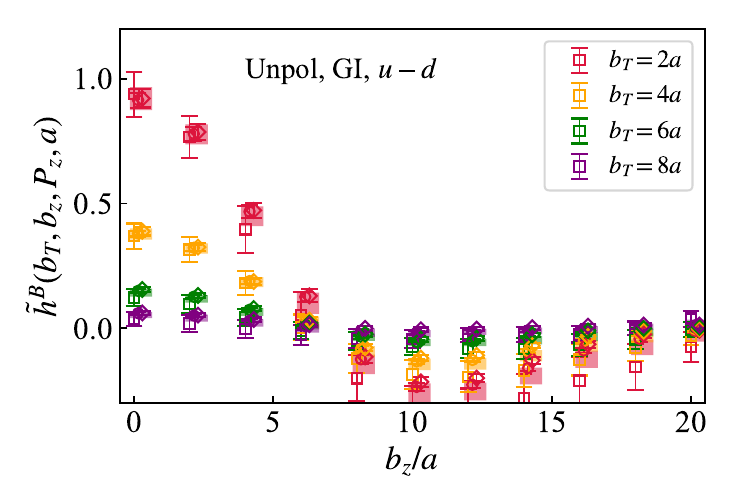}
    \includegraphics[width=0.32\textwidth]{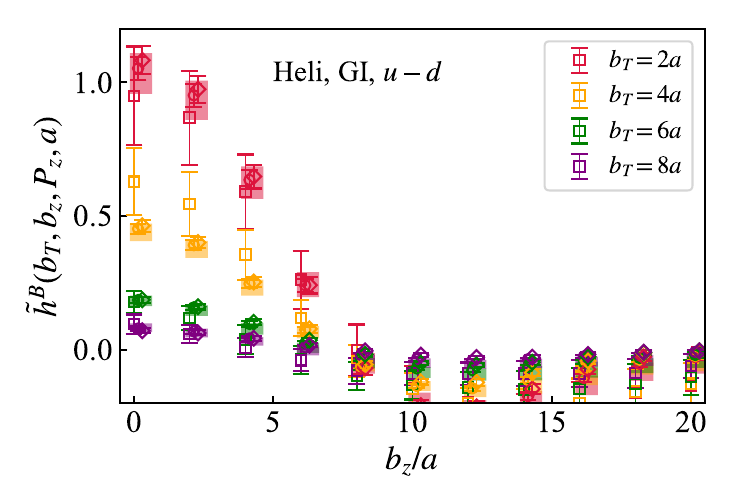}
    \caption{The real-part bare matrix elements for GI quasi-TMD beam functions with nucleon momentum $P_z=1.62$ GeV are shown as functions of $b_z$. The squared, circled and rhombus symbols represent results from two-state, one-state and summation fit respectively. The horizontal error bands are the averaged results obtained from all three methods. From left to right, we show examples for iso-scalar unpolarized, iso-vector unpolarized, and iso-vector helicity cases respectively.\label{fig:bmGI}}
\end{figure}

The GI quasi-TMD beam functions share the same quark propagators as the CG case but require staple-shaped Wilson lines. Therefore, we also computed the GI quasi-TMD beam functions during the contraction, enabling comparisons between GI and CG approaches. We calculated GI quasi-TMD beam function matrix elements with length of the legs of the stable-shaped Wilson link $\eta=12a$, following the same setup as in \refcite{Bollweg:2024zet}. We employed Wilson flow~\cite{Luscher:2010iy}, with a flow time $t_F=2.0$ to suppress the ultraviolet (UV) fluctuations and improve signal quality.

The bare matrix elements for the GI case are also shown in \fig{bmGI}. Compared to the CG case, the GI bare matrix elements decay significantly faster as $b_T$ increases due to the linear divergence from the Wilson line. For instance, at $b_T = 8a$, the GI matrix elements are nearly zero, whereas the CG matrix elements still exhibit sizable amplitudes.

In the main text, we presented the renormalized matrix elements for the CG quasi-TMD beam functions. Here, in the upper panels of \fig{rnmxGI}, we show the examples for the conventional GI approach. Due to artifacts associated with the Wilson line, the signal for GI matrix elements decays rapidly as a function of $b_T$. As a result, reliable data can only be obtained for transverse separations up to approximately $b_T \lesssim 5a$. For the extrapolation and FT, we follow the same strategy as the CG case. In the lower panels of \fig{rnmxGI}, we show the corresponding $x$-dependent quasi-TMD beam functions after FT.

In \fig{cmpCGGI}, we compare the TMD ratios obtained from the CG and GI approaches. The left panel shows $R^{u-d}_{g_{1L}/f_1}(x,b_T)$, the ratio of isovector helicity to isovector unpolarized TMDPDFs, while the right panel displays $R_{f_1}^{u/d}(x,b_T)$, the ratio of up to down quark unpolarized TMDPDFs. The data points represent results from the CG approach, and the horizontal error bands correspond to results from the GI approach. As shown, $R^{u-d}_{g_{1L}/f_1}(x,b_T)$ from both approaches are consistent within uncertainties, indicating that power corrections—arising from finite momentum effects and lattice artifacts—are minimal at this level of precision. Reasonable agreement is also observed for $R_{f_1}^{u/d}(x,b_T)$, although a mild tension within $2\sigma$ appears, likely due to underestimated systematic uncertainties, such as contributions from disconnected diagrams and residual power corrections. Notably, the CG results remain stable up to $b_T \approx 1$~fm, whereas the corresponding GI results decay rapidly. This stability underscores the advantage of the CG methodology, particularly for probing larger transverse separations where non-perturbative effects become significant.

\begin{figure}[th!]
    \centering
    \includegraphics[width=0.32\textwidth]{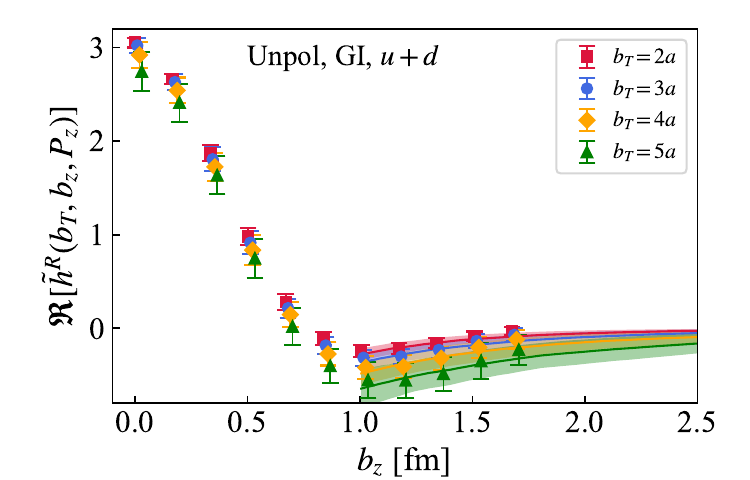}
    \includegraphics[width=0.32\textwidth]{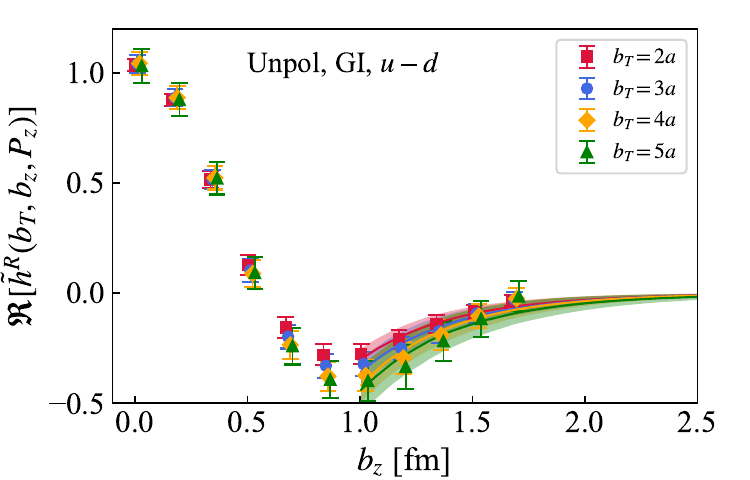}
    \includegraphics[width=0.32\textwidth]{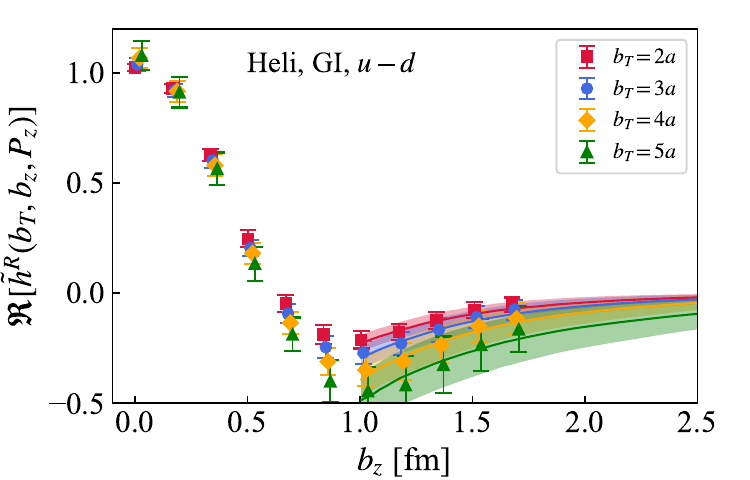}
    \includegraphics[width=0.32\textwidth]{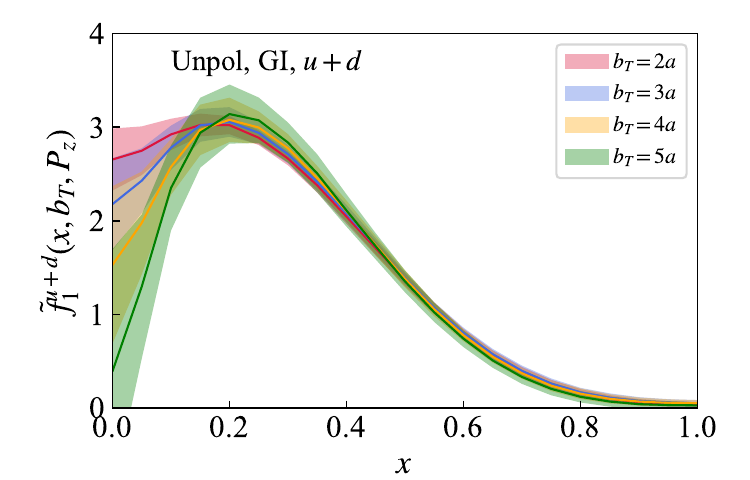}
    \includegraphics[width=0.32\textwidth]{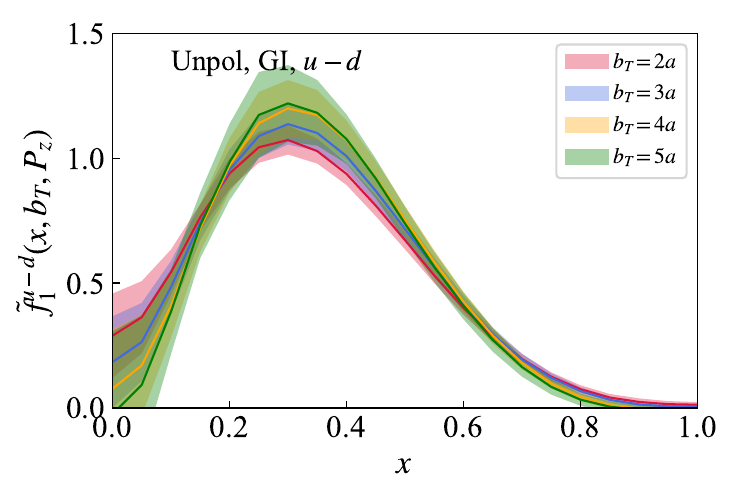}
    \includegraphics[width=0.32\textwidth]{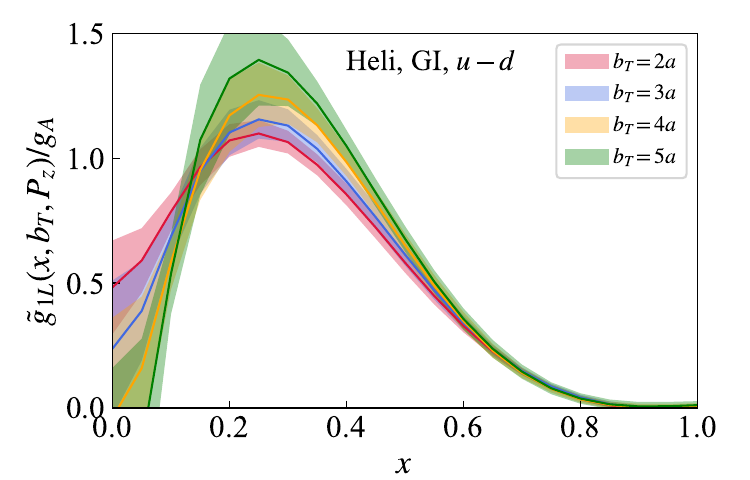}
    \caption{Upper panel: the real-part renormalized matrix elements for GI quasi-TMD beam functions with nucleon momentum $P_z=1.62$ GeV are shown as functions of $b_z$. The bands overlapping with the matrix elements represent extrapolations obtained from $b_z^L=12a$. Lower panel: the corresponding $x$-dependent quasi-TMD beam functions after FT are shown. From left to right, we show examples for iso-scalar unpolarized, iso-vector unpolarized, and iso-vector helicity cases respectively. \label{fig:rnmxGI}}
\end{figure}

\begin{figure}[th!]
    \centering
    \includegraphics[width=0.4\textwidth]{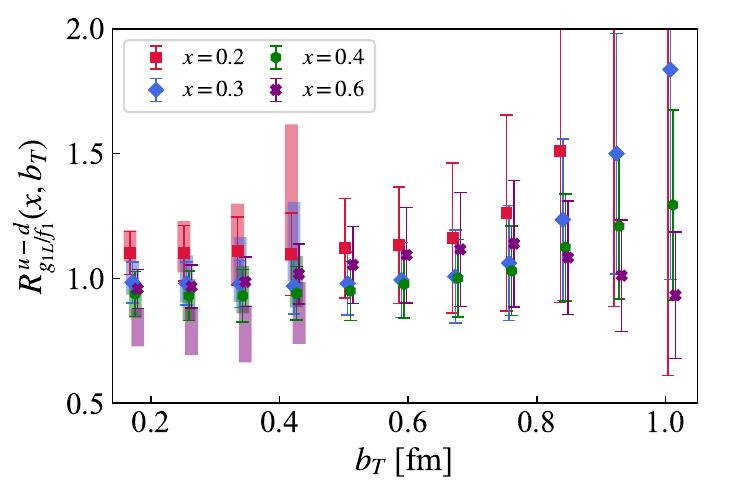}
    \includegraphics[width=0.4\textwidth]{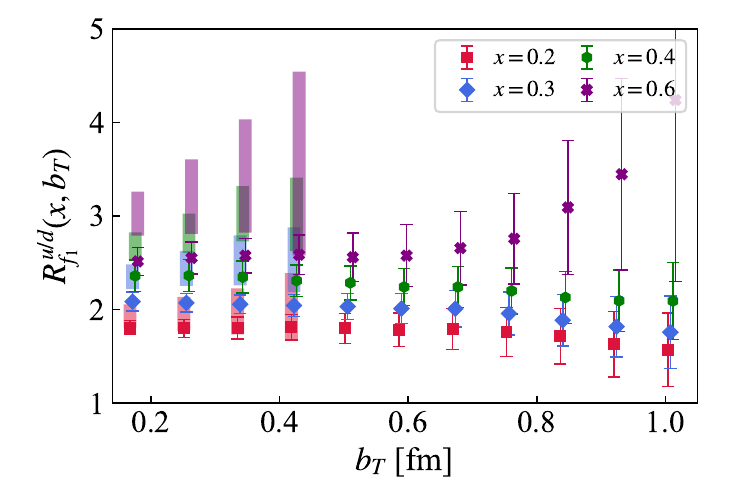}
    \caption{Ratios of TMDPDFs as functions of $b_T$ at various values of $x$ are shown. The left panel displays $R^{u-d}_{g_{1L}/f_1}(x,b_T)$, the ratio of isovector helicity to isovector unpolarized TMDPDFs, while the right panel shows $R_{f_1}^{u/d}(x,b_T)$, the ratio of up to down quark unpolarized TMDPDFs. The data points represent results from the CG approach, while the horizontal error bands correspond to results from the GI approach. \label{fig:cmpCGGI}}
\end{figure}

\end{widetext}
%%%%%%%%%%%%%%%%%%%%%%%%%%%%%%%%%%%%%%%%%%%%%%%%%%%%%%%%%%%%%%%%%%%%%%
\bibliographystyle{apsrev4-2.bst}
\bibliography{main.bbl}
%%%%%%%%%%%%%%%%%%%%%%%%%%%%%%%%%%%%%%%%%%%%%%%%%%%%%%%%%%%%%%%%%%%%%%
\end{document}